\documentclass[a4paper,11pt]{article}
\pdfoutput=1
\usepackage{amsmath,amssymb,xymtex,units}

\usepackage{jinstpub} 

\usepackage{xcolor}
\usepackage{graphicx,epstopdf}
\usepackage{dcolumn}
\usepackage{bm}

\begin{document}
\title{The conceptual design of the miniBeBe detector proposed for NICA-MPD}
\author[a]{Ram\'on Acevedo Kado}
\author[a]{Mauricio Alvarado Hern\'andez}
\author[a,b]{Alejandro Ayala}
\author[c]{Marco Alberto Ayala Torres}
\author[a]{Wolfgang Bietenholz}
\author[d]{Dario Chaires}
\author[a]{Eleazar Cuautle}
\author[e]{Isabel Dom\'inguez}
\author[f]{Alejandro Guirado}
\author[e]{Ivonne Maldonado}
\author[g]{Julio Maldonado}
\author[h]{Eduardo Moreno-Barbosa}
\author[e]{P. A. Nieto-Mar\'in}
\author[a]{Miguel Enrique Pati\~no Salazar}
\author[d]{Lucio Rebolledo}
\author[h,1]{Mario Rodr\'iguez-Cahuantzi\note{Corresponding author.}}
\author[i]{D. Rodr\'iguez-Figueroa}
\author[h]{Valeria Z. Reyna-Ortiz}
\author[h]{Guillermo Tejeda-Mu\~noz}
\author[d]{Mar\'ia Elena Tejeda-Yeomans}
\author[f]{Luis Valenzuela-C\'azares}
\author[h,j]{C. H. Zepeda Fern\'andez}

\affiliation[a]{ Instituto de Ciencias Nucleares, Universidad Nacional Aut\'onoma de M\'exico, Apartado Postal 70-543, CdMx 04510, Mexico} 
\affiliation[b]{Centre for Theoretical and Mathematical Physics, and Department of Physics, University of Cape Town, Rondebosch 7700, South Africa}
\affiliation[c]{Centro de Investigaci\'on y Estudios Avanzados del IPN, Apartado Postal 14-740, CdMx 07000, Mexico} 
\affiliation[d]{Facultad de Ciencias - CUICBAS, Universidad de Colima, Bernal D\'iaz del Castillo No.\ 340, Col.\ Villas San Sebasti\'an, 28045 Colima, Mexico}
\affiliation[e]{Facultad de Ciencias F\'isico-Matem\'aticas, Universidad Aut\'onoma de Sinaloa, Avenida de las Am\'ericas y Boulevard Universitarios, Ciudad Universitaria, C.P.\ 80000, Culiac\'an, Sinaloa, Mexico} 
\affiliation[f]{Departamento de Investigaci\'on en F\'isica, Universidad de Sonora, Boulevard Luis Encinas J.\ y Rosales, Colonia Centro, Hermosillo, Sonora 83000, Mexico} 
\affiliation[g]{Facultad de Inform\'atica, Universidad Aut\'onoma de Sinaloa, Avenida de las Am\'ericas y Boulevard Universitarios, Ciudad Universitaria, C.P. 80000, Culiac\'an, Sinaloa, Mexico}
\affiliation[h]{Facultad de Ciencias F\'isico Matem\'aticas, Benem\'erita Universidad Aut\'onoma de Puebla, Av.\ San Claudio y 18 Sur, Edif.\ EMA3-231, Ciudad Universitaria 72570, Puebla, Mexico}
\affiliation[i]{Universidad de las Am\'ericas, Puebla, Mexico\\ Ex-Hacienda Sta.\ Catarina M\'artir San Andr\'es Cholula, Puebla C.P.\ 72820, Mexico}
\affiliation[j]{C\'atedra CONACyT, CdMx 03940, Mexico}
\emailAdd{mario.rodriguez@correo.buap.mx}

\date{\today}

\abstract{ \vspace*{1mm}
We present the conceptual design for the miniBeBe detector proposed to be installed as a level-0 trigger for the TOF of the NICA-MPD. We discuss the design and the geometrical array of its sensitive parts, the read-out electronics as well as the mechanical support that is envisioned. We also present simulation results for p + p and Bi + Bi collisions to study its capabilities as a function of multiplicity both as a level-0 trigger for the TOF, as well as to serve as a beam-gas interaction veto and to locate the beam-beam interaction vertex.}

\maketitle

\section{\label{sec:intro}Introduction}


The Multipurpose Detector (MPD) is an experiment designed to explore deep into the baryon rich region of the QCD phase diagram by means of colliding heavy nuclei at $\sqrt{s_{NN}}=4-11$ GeV~\cite{MPDgeneral}. The detector is currently being constructed at the Nuclotron-based Ion Collider fAcility (NICA) complex of the Joint Institute for Nuclear Research (JINR). MPD's basic design consists of a central barrel organized in a shell-like structure surrounding the interaction point whose purpose is to reconstruct the traces of both charged and neutral particles in the pseudorapidity range $|\eta|\leq 1.2$. Two end caps will also be placed to detect particles with larger pseudorapidity. The central barrel consists of particle trackers and, during an initial stage, the ones expected to be operating are the Time Projection Chamber (TPC) and the Time of Flight (TOF) systems.

MiniBeBe is a detector designed to provide a wake-up trigger signal for events ranging from low to high multiplicities, for the TOF. The detector name stems from the acronym of \lq\lq Beam-Beam" counter. Given that its dimensions make it to be overall small, the name has been supplemented with the prefix \lq\lq mini".

In order to reliably separate pions, kaons and protons in a wide range of momenta, the TOF is expected to have an overall time resolution better than 100 ps. This requires the trigger signal to be optimized. The nominal MPD element designed to provide this trigger is the Fast Forward Detector (FFD)~\cite{FFD}, which --- in simulations --- has proven to be very efficient for central and semi-central nucleus + nucleus (A + A) collisions, although its efficiency decreases below 50\% for multiplicity events with less than 25 particles. To improve the trigger, the miniBeBe is required to be efficient for low multiplicity p + p, p + A and A + A events as well as to have a fast response. Furthermore, as we show in this work, when the trigger signals from miniBeBe and BeBe~\cite{BeBe} are combined, the trigger efficiency attains about 80\%.




To produce a trigger signal for the TOF, one can envision placing a fast, low-cost detector surrounding the interaction point. In order to avoid distorting the properties of particles produced in the collisions to be studied, this detector is required to have also a low material budget. Low-cost fast detectors are nowadays based on the combined use of thin and small transverse area plastic scintillator cells coupled to Silicon Photo Multipliers (SiPMs). As shown in this work, the size, thickness and number of SiPMs of a cell can be optimized to achieve a fast response signal of order 20-30 ps. If this fast response is combined with fast read-out electronics with a response time of about 20 ps, it is then conceivable that the designed detector can serve as a good TOF trigger, provided it proves to be efficient for low multiplicity events. 

In fact, as we also show in this work, the read-out electronics can achieve
resolution times below 10 ps with a band width in the 10 GHz range with rise and fall times of order 20 ps. The low material budget criterion is met provided the mechanical support is made of light, yet firm, material. In this work we also provide studies showing that the design can meet this requirement and we present Monte Carlo studies showing that the detector does work for low multiplicity events and that its capabilities could even be improved when its longitudinal dimensions are increased, its transverse dimensions are reduced and the number of sensitive elements is increased. As for any evolving design, some points like the characteristics of the cooling system, the cabling, the position of the read-out electronics, etc., are being also actively pursued and their discussion is being reserved for a more detailed Technical Design Report.



\begin{figure*}[hbt!]
\includegraphics[trim=0.2cm 0.3cm 0.3cm 0.5cm ,clip,scale=0.3]{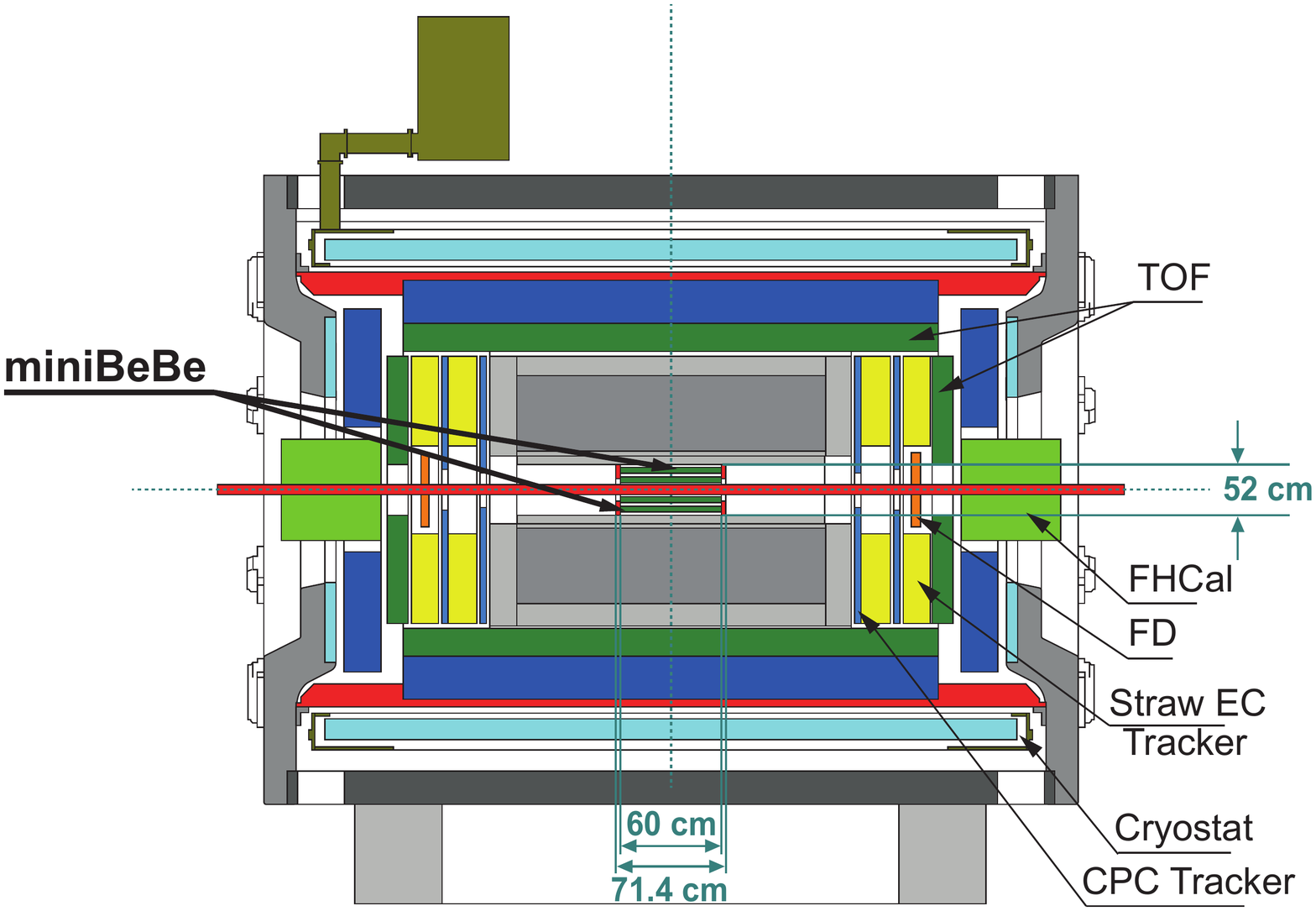}\includegraphics[scale=0.18]{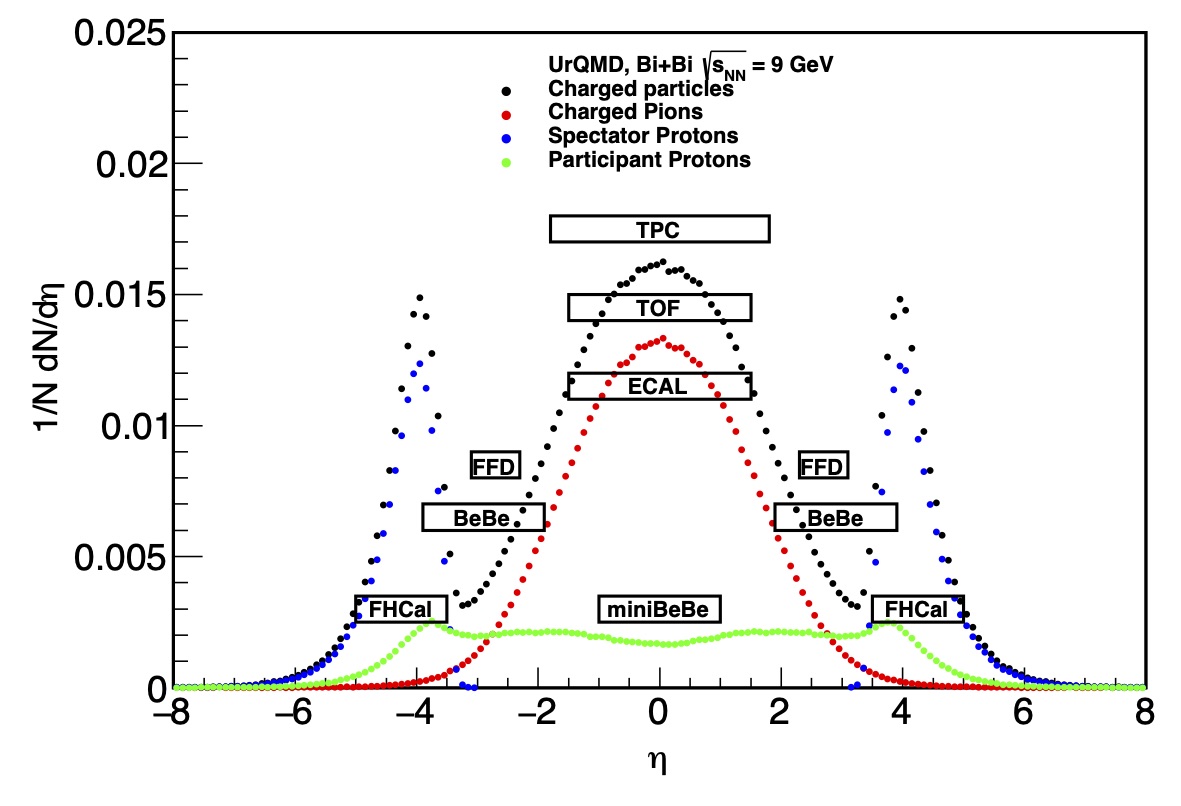}
    \caption{(Left) Schematic representation of the planned location of the miniBeBe detector inside the other MPD components. (Right) Pseudorapidity coverage of the miniBeBe detector (represented by width of the named box) compared to the nominal pseudorapidity coverage of the rest of the MPD components: Time Projection Chamber (TPC), Time of Flight (TOF), Electromagnetic Calorimeter (ECAL), Fast Forward Detector (FFD), Beam-Beam Monitor Detector (BeBe), Forward Hadron Calorimeter (FHCal) and miniBeBe. The curves correspond to the pseudorapidity distributions for charged particles (black), charged pions (red), spectator protons (blue) and participant protons (green) computed for a UrQMD sample of Minimum Bias events of Bi + Bi collisions at $\sqrt{s_{NN}}=9$ GeV}.
    \label{fig:mpd}
\end{figure*}
 
Here we describe the concept for the miniBeBe {\it baseline} design and report on the progress of the construction of its parts, including the array of sensitive elements, the read-out electronics and the mechanical support. We also present results from simulations to explore its performance as a trigger under different multiplicity environments.

The work is organized as follows: in Sec.~\ref{sec:secII} we describe the overall detector concept. In Sec.~\ref{secIII} we present the details of the front-end electronics and in Sec.~\ref{secIV} the mechanical structure designed to support the sensitive elements and the electronics. In Sec.~\ref{secV} we discuss the material budget introduced by the miniBeBe in terms of the energy loss of charged particles that pass through its components. In Sec.~\ref{secVI} we present simulation results to estimate the intrinsic time resolution of a basic cell. Sections~\ref{secVII} and~\ref{secVIII} show our results of simulations to study the time resolution and trigger capabilities of the miniBeBe using p + p and A + A collisions. Finally, we summarize and conclude in Sec.~\ref{concl}.

\section{Baseline design}\label{sec:secII}

In order to achieve the fast trigger signal and low material budget requirements, we propose a baseline geometry for the miniBeBe which consists of 16 strips, each one of length 600 mm. The strips are made of arrays consisting of 20 squared plastic scintillator cells with dimensions $20 \times 20 \times 3\  \mathrm{mm}^3$. The remaining 200 mm correspond to the total length obtained by adding the space (10 mm) between adjacent cells and is occupied by the support structure and by the electronics card. 

There are 4 SiPMs coupled to each cell. The strips are supported by a cylindrical, lightweight shell, whose inner and outer radii are 220 and 260 mm, respectively, measured from the symmetry axis of the beam pipe. The  radii might still vary to possibly improve the detector capabilities. The design has the advantage of being modular in the sense that the number and length of the strips can be adjusted to accommodate a longer and/or smaller radius detector. For the purpose of this work, we concentrate on the description of the baseline geometry.

In order to optimize the design, we have performed simulations for p + p and A + A collisions, using standard Monte Carlo generators such as PYTHIA 8, UrQMD and PHSD. From these studies we found that it is possible to achieve a fast trigger covering the range $|\eta|<1.01$. The planned location and rapidity coverage of the miniBeBe, compared to the rest of the MPD components, is shown in Fig.~\ref{fig:mpd}.

In total, the miniBeBe is made of 320 squared plastic scintillator cells and 1,280 SiPMs covering an effective sensitive area of 128,000 $\mathrm{mm}^2$. Since the plastic cells are mounted over the surface defined by the inner radius, the corresponding sensitive area represents $15.43 \%$ of the total cylinder area.

\begin{figure*}[!htb]
\centering
\includegraphics[scale=0.2]{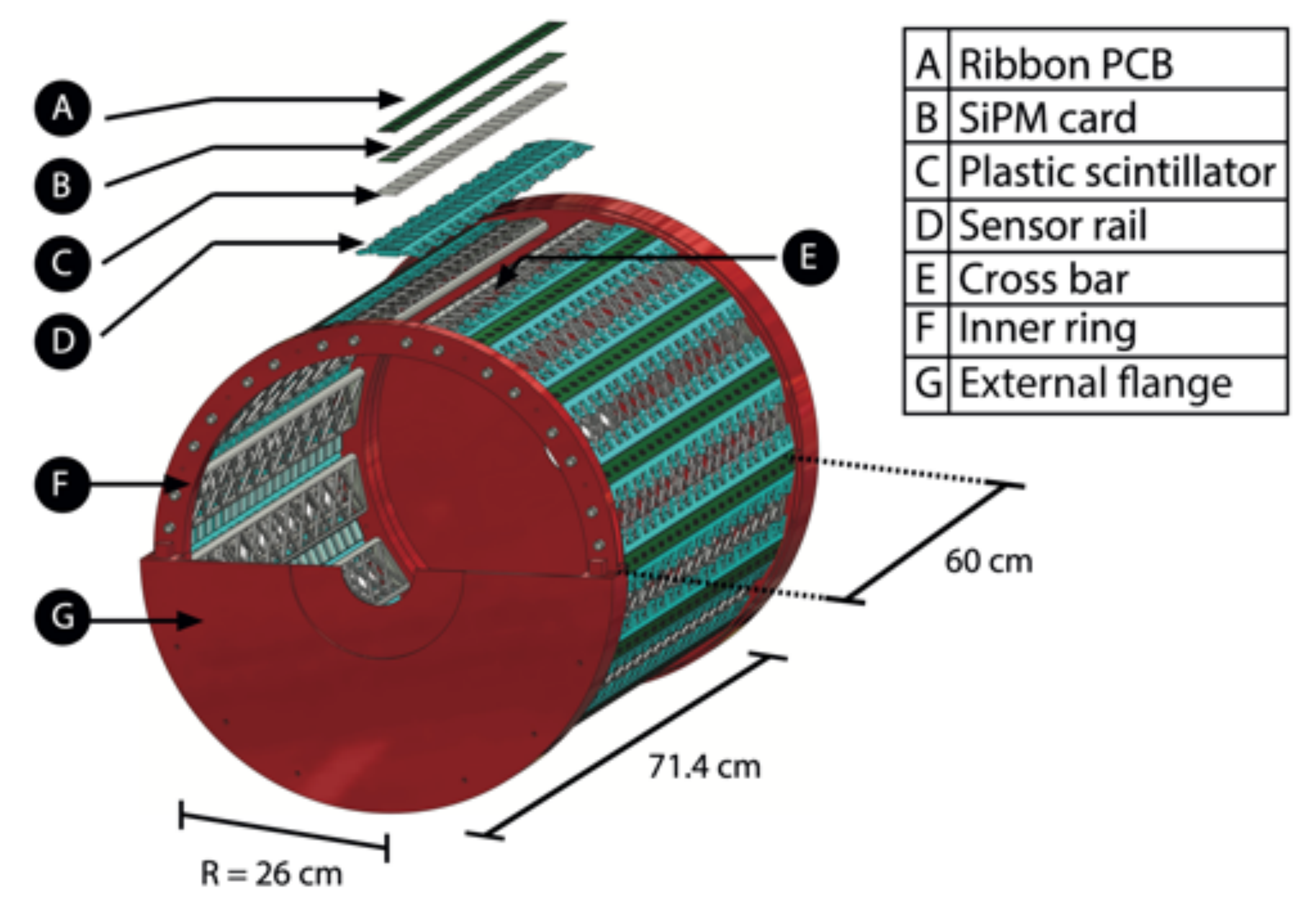}
\caption{Illustration of the miniBeBe detector. The structure holds sixteen 600 mm long strips mounted on a cylinder, with inner and outer radii of 220 and 260 mm, respectively, placed around the beam pipe. 
Each strip consists of 20 squared plastic scintillators with dimensions $20 \times 20 \times 3\  \mathrm{mm}^3$, with four SiPMs coupled to each cell.}
\label{minibebe}
\end{figure*}

To strengthen the mechanical integrity of the support, two external flanges are added as end caps of the cylinder, each having a 57 mm width. The concept and overall size is depicted in Fig.~\ref{minibebe}.

Each plastic cell is made of BC404 plastic scintillator~\cite{BC404}. This is a commonly used plastic scintillator for fast counting applications~\cite{fasta,fasta1}. 
It has a base of polyvinyltoluene with a refractive index of 1.58, a density of 1.023 g/cm$^3$ and a light decay constant of 1.8 ns. This kind of plastic scintillator can be used in vacuum environments. The wavelength of maximum emission is 408 nm. Its softening point is at $70^{\circ}\,$C. For the BC404, the average number of photons produced by a fast charged particle going through the cell is 551. For completeness, we have also considered cells made of BC422 plastic scintillator, which produces in average 509 photons as the response to a fast moving charged particle. We chose the former since its light yield is larger. The SiPMs make use of a recent technology for silicon semiconductors. Unlike previous semiconductor-based models, these devices have the ability to resolve even a single photon.  The selected model for the miniBeBE detector is the MicroFC-60035 SensL SiPM with dimensions 6$\times$6 mm$^2$ manufactured by SensL Technologies, Ltd., with a cell length of 35 $\mu$m, for a total of 18,980 cells distributed over its $6 \times 6$ mm$^2$ surface. 
SensL SiPMs are produced having a {\it fast} output pin. As shown in~\cite{Zepeda2020}, this output corresponds to the time derivative of the standard output signal. Its rise time is 1 ns and its pulse width is 3.2 ns. The maximum of the Photon Detection Efficiency (PDE) is reached for a wavelength of 420 nm, ranging from 31\% to 41\% for an over-voltage of 2.5 V and 5 V, respectively, and corresponds to a gain of $3\times 10^6$~\cite{SiPM}.

Starting from this baseline design we now concentrate on the description of the front-end electronics, optimal cell occupancy, best performance as trigger, minimal material budget and expected improvements for the miniBeBe to serve as a TOF trigger.

\section{Front-End Electronics}\label{secIII}

The main goal for the design and implementation of the front-end electronics is to generate trigger pulses for the TOF, based on the detection of fast moving particles.
For the output, only the fast signal is used since it has a better timing response compared with the standard signal. As recommended in Ref.~\cite{SensL2013}, a voltage higher than the breaking voltage $V_{\rm br}$, given by $V_{\rm br}+5$ V, was used to maximize the SiPM PDE. The fast signal must have a load resistance in order to generate a current path to the reference ground. Hence, a $50\ \Omega$ resistor is used as an output load with an analog output signal.

The time resolution for different SiPMs attached to several scintillating materials has been thoroughly studied~\cite{Gundaker2020,Chytka2019,Betancourt2020}. In addition, Ref.~\cite{Dolinsky2013} shows that for both fast and standard output signals, SensL SiPMs similar to the ones proposed to be used in the miniBeBe design, have a time resolution of order 10 ps, when using front-end electronics in the 1 GHz band. This time resolution can be improved --- or at least maintained --- when using front-end electronics in the 10 GHz band, with rise and fall times of order 20 ps~\cite{HMC674}, such as the one we are currently pursuing.

\begin{figure*}[!htb]
\centering
\includegraphics[scale=0.4]{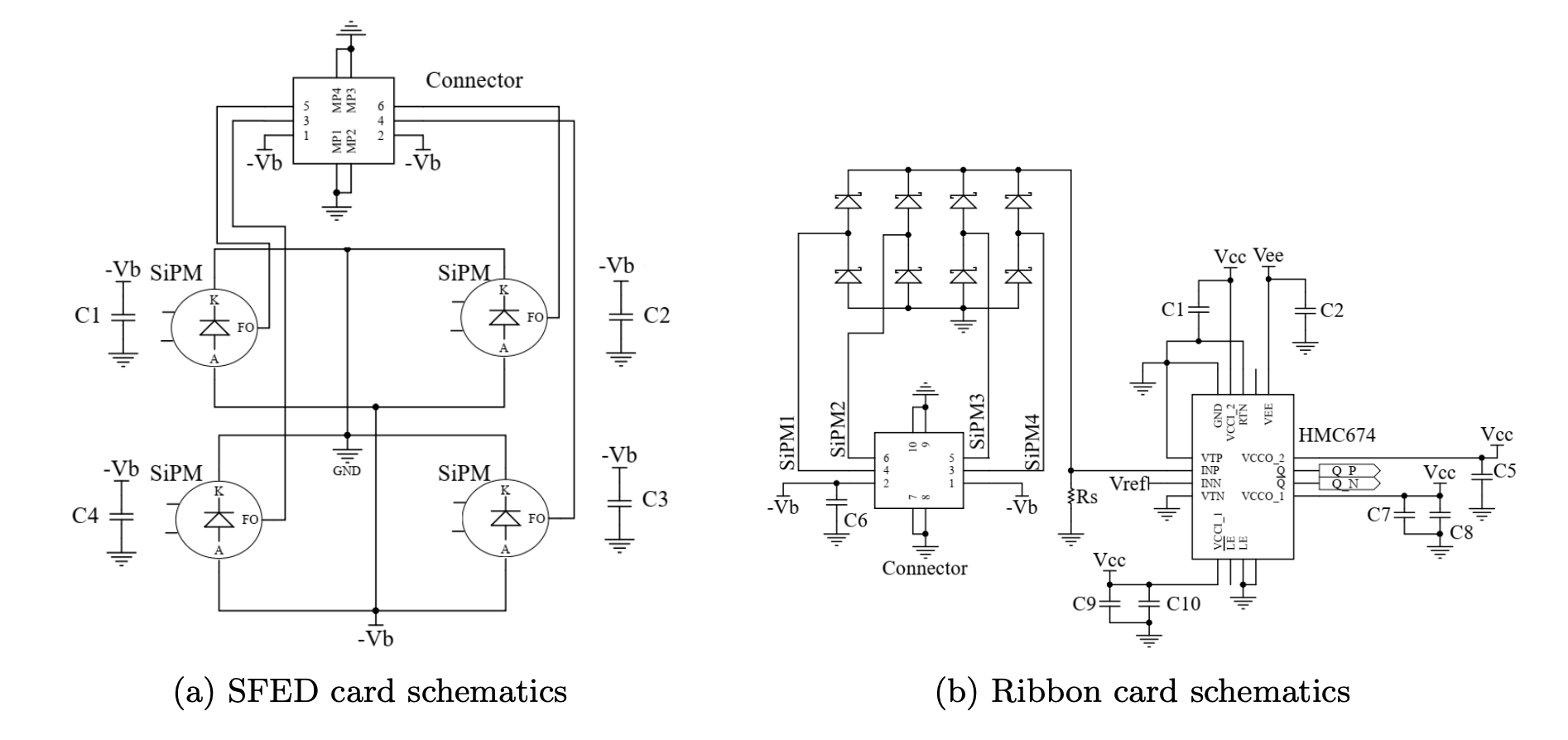}
\caption{Front-end electronics for (a) the SFED card and (b) a single channel trigger in the ribbon card.}
\label{fe_SiPM}
\end{figure*}

Four SiPMs are attached to each of the plastic cells which are in turn placed over a printed circuit board. The schematic design is illustrated in Fig.~\ref{fe_SiPM}(a), where we show how decoupling capacitors and a nano connector are used for this PCB. The analog signals come from the fast output of each SiPM. As shown in Ref.~\cite{Zepeda2020}, the charge information can be acquired from the analog fast output signal.

The PCB is designed on a Flame Retardant 4 (FR-4) material (that complies with the NEMA UL94V standards) with a dielectric constant of 4.34 at 1 GHz.

The output analog signal is transformed into a digital differential signal by using the analog comparator HMC674~\cite{HMC674} with an input bandwidth of $9.3$ GHz, $85$ ps of propagation delay, input minimum pulse width of $60$ ps and output rise time of $24$ ps. The logical voltage levels for this signal correspond to the Reduced Swing Positive Emitter-Coupled Logic (RSPECL) standard, described in Table~\ref{tab:RSPECL}~\cite{HMC674}.

\begin{figure}[!hbt]
  \centering
 \includegraphics[scale=0.1]{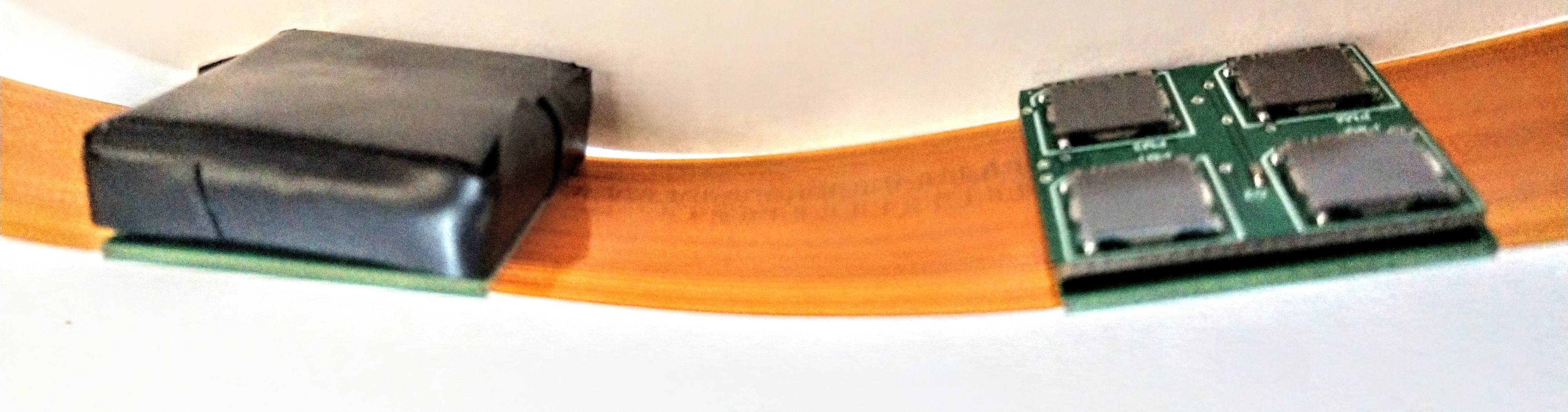}
\caption{General front-end PCB prototype. The picture shows the SFED card attached to the ribbon card. On the left, the SFED card is wrapped in polyester mylar tape. On the right, the SFED card is unwrapped and the SiPMs distribution is visible}
\label{slidembb}
\end{figure}

Figure~\ref{fe_SiPM}(b) shows a schematic representation of the electronics for the connection between the SFED card and the back plane. The latter is referred to as the \lq\lq ribbon card". The logical \lq\lq OR" for the fast output signals is implemented by means of an array of Schottky diodes. The resulting analog signal passes trough the analog comparator for discrimination. The digital signal is obtained with a pulse width proportional to the analog pulse width of the fast output signal. The analog comparator used for this initial prototype is the HMC674 with an input bandwidth of $9.3$ GHz,  85 ps of propagation delay, input minimum pulse width of 60 ps and output rise time of 24 ps. The logical voltage levels for this signal correspond to the Reduced Swing Positive Emitter-Coupled Logic (RSPECL) standard, described in Table~\ref{tab:RSPECL} \cite{HMC674}.
It is important to notice that Fig.~\ref{fe_SiPM} shows just a single channel for the backplane ribbon card. Given the baseline dimensions of the entire miniBeBe, 20 such channels are considered for the design. In order to avoid an excess of material budget, the connector at the end of each ribbon card will be placed away from the center of the detector. Two ribbon cards, each with a length of 30 cm, will be used to cover the total length of 60 cm, with each card accommodating 10 plastic cells, 10 differential pair RSPECL signals and each consuming the bias voltage and using two symmetrical power sources of $\pm$ 3.3 V for the discriminator circuits on each channel. The ribbon card is made of a rigid-flex material. The analog to digital conversion is performed in the rigid part. Once again, in order to avoid cabling and material budget excess, each pair of trigger signals is sent through the ribbon cards. The rigid part of the PCB is designed on a FR4 material with a dielectric constant of 4.34 at 1~GHz, while the flexible part is designed of polymide with thickness of 0.05 mm, 1 oz of copper and permittivity of 3.78 at 1~GHz. An actual picture for SFED and ribbon PCB cards is shown in Fig.~\ref{slidembb}.

\begin{table}[b]
	\centering
	\begin{tabular}{c|c|c|c|c}
		\hline\noalign{\smallskip}
		Parameter             & Min.  & Typ. & Max. & Units \\
		\noalign{\smallskip}\hline
		High level             & 1.03  & 1.09 & 1.14  & V\\
		Low level              & 0.65  & 0.71 & 0.81  & V\\
		Differential swing     & 440   & 760  & 980  & mV (p-p)\\
		\hline 		
	\end{tabular}
		\caption{Voltage levels for the RSPECL standard.}
	\label{tab:RSPECL}
\end{table}

\begin{table}[b]
	\centering
	\begin{tabular}{c|c|c|c|c}
		\hline\noalign{\smallskip}
		Parameter             				 & Min.  & Typ. & Max. & Units \\
		\noalign{\smallskip}\hline
		1 SFED voltage			    		 & 27.5  & 29 	  & 30	  & V \\
		1 SFED current			    	     & 80  	& 100	& 120  & mA \\
		1 Analog comparator power	         & --   	& 140   & --  	 & mW \\
		1 Analog comparator voltage	         & $-3.3$   & --	  & 3.3   & V  \\
		TRB3 voltage						 & --  	 & 48	 & 980  & V \\
		TRB3 current						 & --	   & 10	    & --		 & A \\
		\hline 		
	\end{tabular}
		\caption{Power supply requirements.}
	\label{tab:PowerSupply}
\end{table}

\begin{figure*}[!hbt]
	\centering
	\includegraphics[width=1\linewidth]{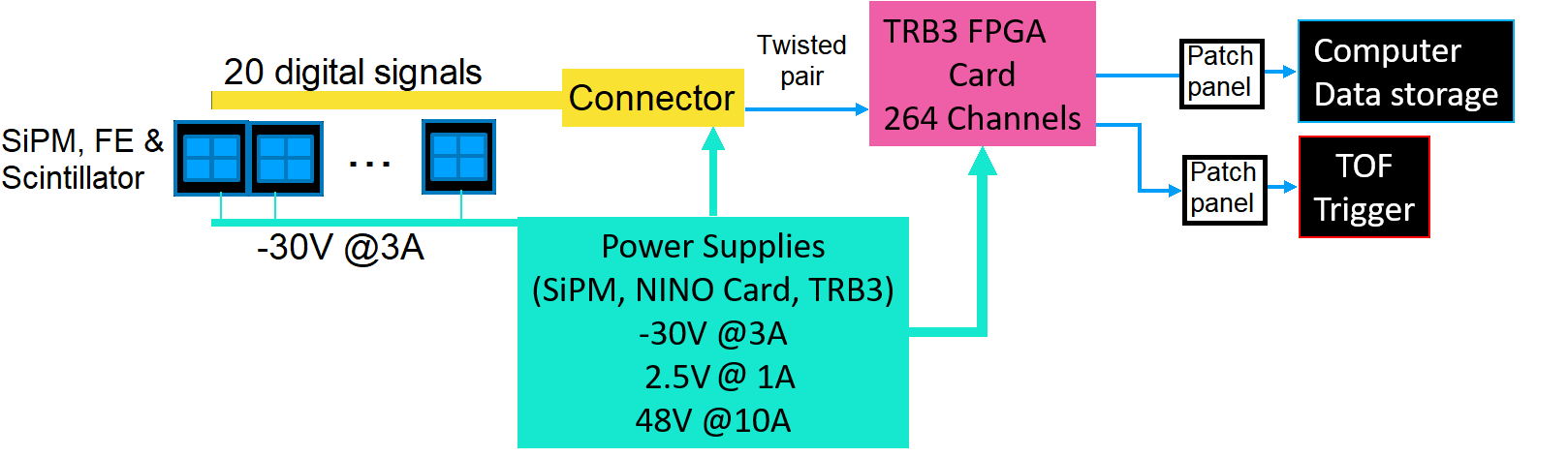}
\caption[]{General schematics of the miniBeBe front-end electronics.}
	\label{fig:feschematics}
\end{figure*}

All the trigger signals are collected using a TRB3 Field Programmable Gate Array (FPGA) card, controlled from a Linux computer to acquire and store up to 264 input channels of information in a data center~\cite{trb3}. Part of the signal processing task will be developed inside the FPGA card. Thus a single trigger for the TOF sensor will be generated inside this FPGA card, achieving the main objective of this front-end. As described by the general schematics shown in Fig.~\ref{fig:feschematics}, a power supply bank is required with low ripple, high pass filtering and good grounding system to avoid interference and noise induction to all the front-end design. The voltage and current requirements are specified in  Table~\ref{tab:PowerSupply}.

At present, the location of the FPGA card is still under consideration. However, we have considered  $3$ m of cabling, from the end of each ribbon card, as a possible means to extract the signal. By using a twisted pair cable, a time delay close of 4~ns per meter is expected~\cite{Bates2017}. Figure~ \ref{fig:feschematics} represents the twisted pair cable as a line, between the nano D connector and the TRB3 card.

\section{Mechanical structure}\label{secIV}

The mechanical structure consists of the main support for the plastic scintillators cells and for the read-out electronics. This is schematically shown in Fig.~\ref{minibebe}.

\begin{figure}[!hbt]
  \centering
\includegraphics[scale=0.36]{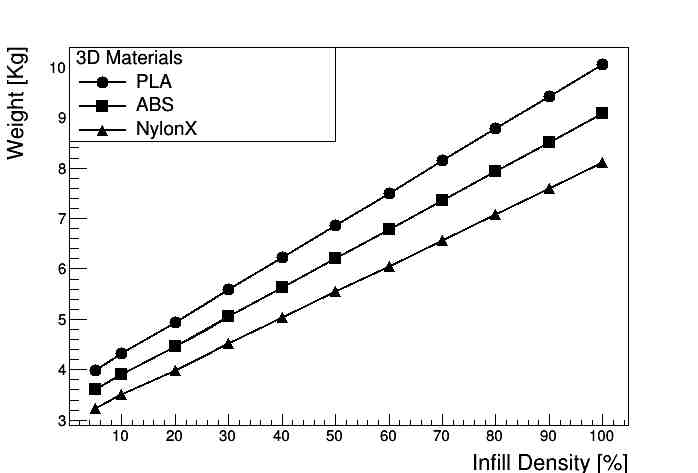}
\caption{Estimated weight as a function of the print density for Polylactic Acid (PLA), Acrylonitrile Butadiene Styrene (ABS) and NylonX (Carbon Fiber Reinforced Nylon) filaments.}
\label{weightable}
\end{figure}

The mechanical structure has been designed accounting for the requirement of a low material budget, which was translated into a lightweight but at the same time firm structure. This design considers an eventual 3D printing consisting of removable pieces allowing to eventually assemble the essential parts and adjust rigidity and precision for the overall structure. Figure~\ref{weightable} shows the estimated weight of the mechanical support as a function of different density percentages of 3D printing materials~\cite{weightableest}.


The structure has been developed having in mind the Plug\&Play concept and the possibility to replace the rails that support the electronics and plastic scintillators at will, without having to disassemble the whole structure.

Deformation simulations of the structure parts were performed using finite element analysis with the {\it Autodesk Inventor} software, to approximate the behavior of the structure under extreme conditions of temperature variations and of differential pressure. Table~\ref{tab:volume} shows the volume corresponding to each of the structure parts, as an indicator for finite element simulations. The design considers the mass of each integrated element within the miniBeBe structure.


\begin{table}[b]
	\centering
	\begin{tabular}{|l|r|}
	\hline
	External flanges & 1373094.2 mm$^3$ \\ 
	\hline
	Inner rings & 556121.1 mm$^3$ \\ 
	\hline
	Cross bars & 115860.1 mm$^3$ \\ 
	\hline
	Sensor rails & 77380.0 mm$^3$ \\ 
	\hline
	Top cover for rails & 44776.0 mm$^3$ \\ 
	\hline
	\end{tabular}
		\caption{Volume of the miniBeBe support structure components used in our simulations.}
	\label{tab:volume}
\end{table}

\begin{figure}[!hbt]
\centering
\includegraphics[scale=0.31]{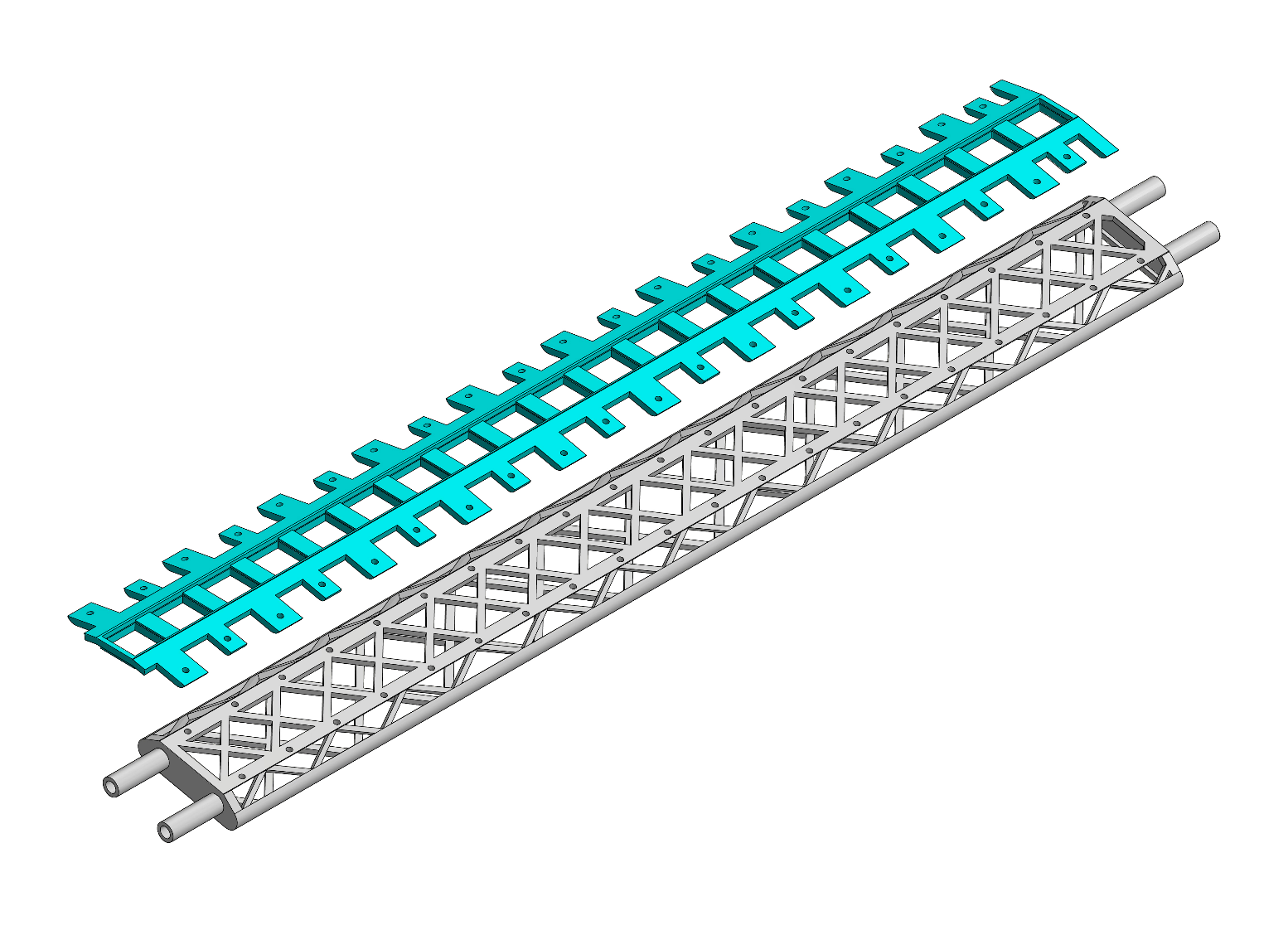}
\caption{Sensor rail to hold 20 plastic scintillators with dimensions $20 \times 20 \times 3$ mm$^{3}$ (top) and cross bar support for the sensor rails (bottom).}
\label{sensorail}
\end{figure}

Figure~\ref{sensorail} shows the schematic representation of a sensor rail (top) that holds the plastic scintillators and the cross bar support (bottom). The rail is designed to hold a strip consisting of 20 cells of dimensions $20 \times 20 \times 3$ mm$^{3}$, each connected to its corresponding read-out electronics. The rails are to be screwed to the cross bars to provide support, rigidity and stability. 

\begin{figure}[!hbt]
\centering
\includegraphics[scale=0.13]{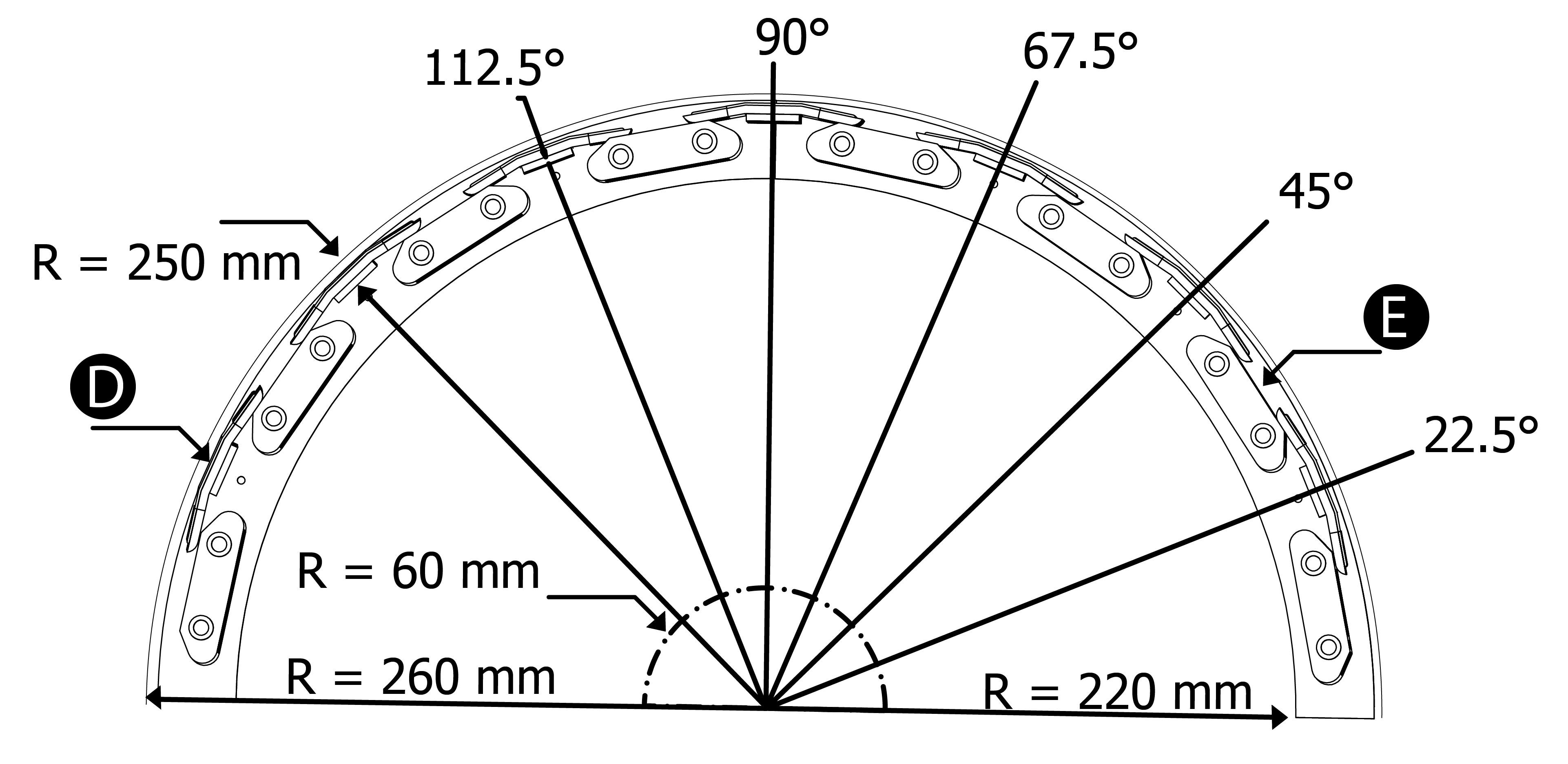}
\caption{Transverse plane view of the cross bar array for the sensor rails. The labels D and E refer to the elements depicted in Fig.~\ref{minibebe}.}
\label{angular}
\end{figure}

The design also considers simulations carried out within the MPDRoot~\cite{MPDRoot} frame for a 16 strips cylindrical geometry. 
The whole structure is designed so that the detector cells are located 250 mm from the beam axis. Each rail is separated by 22.5$^\circ$ in the transverse plane. The support has an external radius of 260 mm and an internal radius of 220 mm. The latter corresponds to the ring that supports the cross bars. The caps have a 60 mm internal radius, so that direct contact with the beam pipe is avoided. The whole cylinder consists of two sections with a semicircular cross section on the transverse plane that can be clamped together around the beam pipe. A schematic representation of one of the cylinder halves, viewed from the transverse plane, is shown in Fig.~\ref{angular}, where the dimensions described above can also be seen.

For tolerance tests and structural alignment of the cylinder, 3D prints were made at a density of 10\% in Polylactic Acid (PLA) and NylonX (nylon reinforced with carbon fiber) in order to obtain a prototype for manufacturing in 100\% carbon fiber using additive manufacturing technologies. 

The structure is designed for easy assembly. Each of the strips is individually assembled over the support rails and then placed on the cross bars to be later screwed together. This makes maintenance and replacement of parts quick and easy. 

\section{ Material budget}\label{secV}

In order to estimate the possible effect on the energy of particles passing through the detector material, we have also performed studies of the energy loss of primary particles (pions and muons) in the range of 5 MeV to 5 GeV. To assess the effect of the different detector materials, we perform the analysis both for the Detector Element (DE) as well as for the blind area (BA). The former consists of the BC404 plastic scintillator to which the SiPMs are attached together with the Polyvinyl Chloride (PVC) where the electronic circuits are printed. The latter is considered to be made of polyacrinolitrile. The simulation studies were made using the Geant4 software. The BA is taken to have a thickness of 6.56~mm 
while the DE thickness is 4~mm. We find that there is no distinction between the energy deposit of the two considered primary particles. For the DE and the BA, the energy deposited per particle is in the range of 0.49 to 0.94~MeV, and 1.35 to 2.58~MeV, respectively. These findings are summarized in Table~\ref{loss}.

To complement these studies, we also compute the percentage of the characteristic radiation length $X0$ introduced by these materials. For the DE and BA, the contribution from the BC404 plastic scintillator cells is the dominant one. Considering that the BC404 plastic has a density of $\rho_{BC404}=1.023$ gr/cm$^3$, its contribution to the radiation length is 0.7\%$X0$. On the other hand, the mechanical structure is envisioned to be made of carbon fiber 40 mm thick. Considering that carbon fiber has a density of $\rho_{CF}=1.93$ gr/cm$^3$, and that our {\it spider web} design for the cross bars takes up about 20\% of the volume, its contribution to the radiation length is of order 0.36\%$X0$. For comparison purposes, the current design of the MPD Internal Tracking System (ITS) considers a total contribution of its material budged of order 0.8\%$X0$~\cite{MPDgeneral}. The optimization of the material budget for the rest of the miniBeBe components (ribbon card, SiPMs card, external flange, overall cross bar structure) is work in progress that will be presented in a more comprehensive technical design report.

\begin{table}[hbt]
\begin{center}
\begin{tabular}{|l|l|l|}
\hline
  IE (GeV) & E$_{\rm loss}$ in DE (MeV) & E$_{\rm loss}$ in BA (MeV)\\
  \hline
  0.05                     &  0.94 $\pm$ \ 0.01              & 2.58 $\pm$ \ 0.23   \\
  \hline
  0.1                       &  0.67 $\pm$\ 0.07              & 1.85 $\pm$\ 0.18  \\
  \hline
  1                          &  0.48 $\pm$\ 0.01              &  1.35 $\pm$\ 0.15  \\
  \hline
  3                          &  0.49 $\pm$\ 0.06              &  1.35 $\pm$\ 0.15 \\
  \hline
  5                         &  0.49 $\pm$\ 0.06              &  1.35 $\pm$\ 0.15 \\
  \hline
\end{tabular}
\end{center}
\caption{Energy loss (E$_{\rm loss}$) of primary particles (pions and muons) with a given Incident Energy in DE and in the BA of the miniBeBe. The energy loss is negligible for the considered range of incident energy and thus we do not expect the material budget will to affect the particle properties while passing through the detector.}
\label{loss}
\end{table}

\section{Geant4 Simulations to estimate the intrinsic time resolution for a basic cell}\label{secVI}
\begin{figure}[!htb] 
\centering
  \includegraphics[width=20pc]{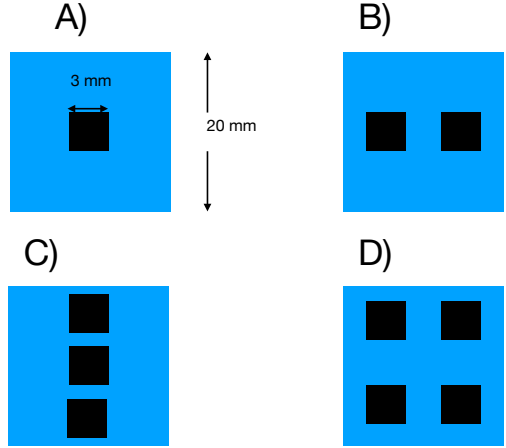}\hspace{2pc}
  \caption{\label{conf}Illustration of the four scorer configurations that we simulated in order to identify the one with the optimal time resolution.}
\end{figure}

In order to study the intrinsic time resolution for the basic elements of the miniBeBe, we performed simulations using Geant4 v10.06p01~\cite{g4}. The configurations we study consist of arrays of one, two, three and
four SiPMs of size $3 \times 3~{\rm mm}^2$ distributed on the surface of the plastic scintillator cells.
The intrinsic time resolution is studied without including the contribution due to the electronic output. The different configurations we consider are depicted in Fig.~\ref{conf}, where the black squares
(scorers) represent the SiPMs. The goal is to
explore the configuration that provides the minimal time resolution. This is carried out considering also two kinds of plastic scintillators: BC404 and BC422.

We simulated 1000 $\pi^+$-mesons, arriving one by one at the cell where they hit the full frontal area, on the opposite face of the one where the scorers are located. The $\pi^+$ are given an average kinetic energy of 0.5~GeV, which corresponds to their typical energy for A + A collisions at NICA energies.
For each event, we recorded the lowest mean of the Landau time-of-flight distribution obtained in any one of the scorers. This time represents the first pulse. 
For the BC404 plastic scintillator, our results imply an intrinsic time resolution of
$7.76 \pm 0.87~{\rm ps}$ and $9.29 \pm 0.67~{\rm ps}$,
for one and four scorers, respectively. 
For the BC422 plastic scintillator we obtained $7.76 \pm 0.87~{\rm ps}$ and
$9.29 \pm 0.75~{\rm ps}$, for one and four scorers, respectively. However, these differences of up to 2~ps are not significant
in light of the fact that the electronics has only a time resolution of about 20~ps \cite{trb3}. In this sense, the time resolution is equivalent for all scorer configurations and both scintillator materials. Notice that these results refer only to the intrinsic time resolution of the plastic scintillators and do not account for the SiPMs PDE. Nevertheless, a quick estimate can be made to include the SiPM PDE. Considering the BC404, an average of 551 photons reach the SiPM surface. Therefore, using a 41\% PDE, 226 photons are expected to be detected by a single SiPM. For the BC422, this number is, correspondingly, 208.

\begin{figure}[!htb]
\centering
  \includegraphics[width=21pc]{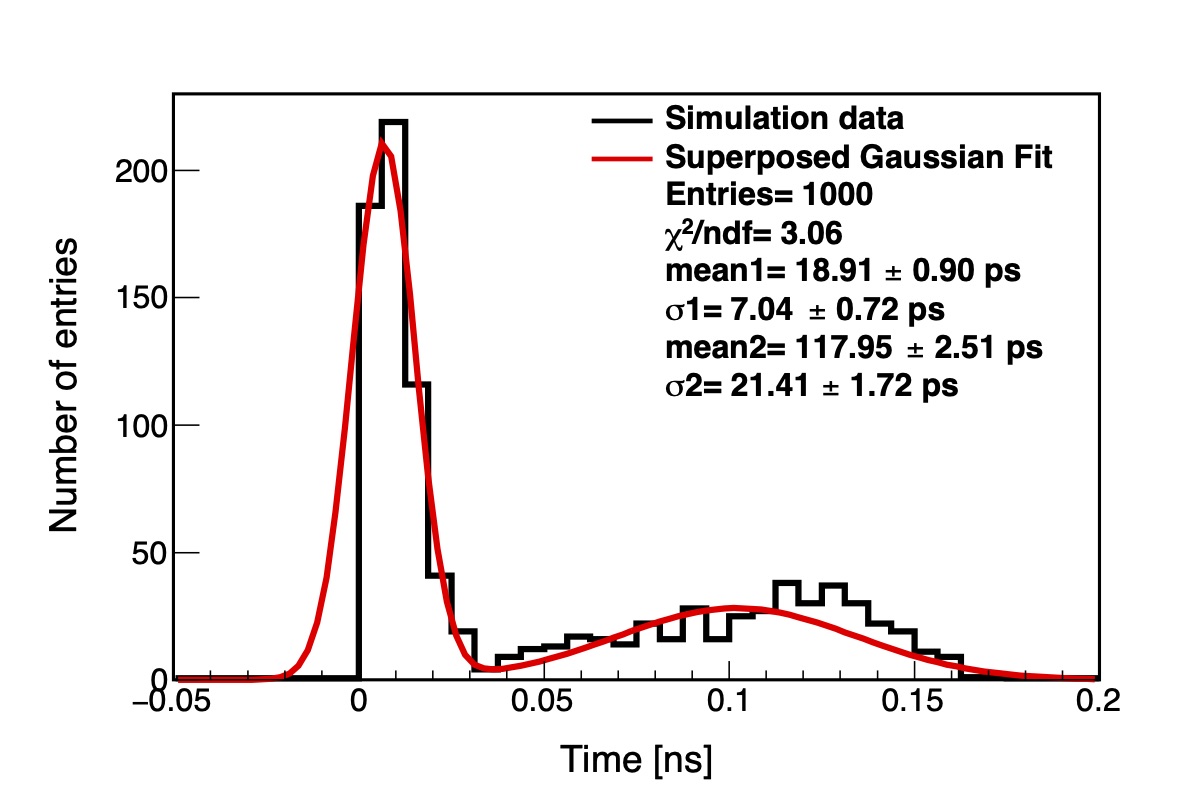}\hspace{2pc}
  \caption{\label{gaus}Time of flight distribution for photons produced by the plastic scintillator. We show the results for configuration D in Fig.~\ref{conf}.} 
\end{figure}

\begin{figure}[!htb]
\centering
  \includegraphics[width=21pc]{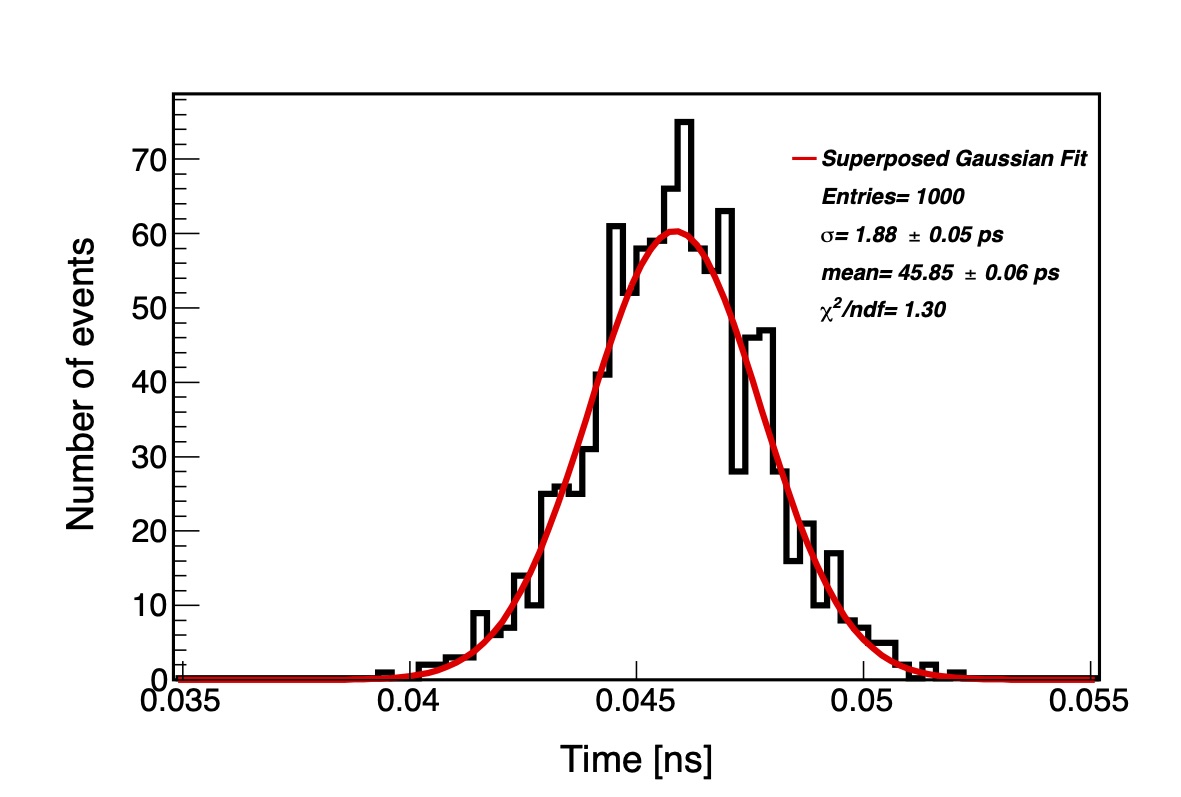}\hspace{2pc}
  \caption{\label{top}Time-of-flight distribution for photons produced by the plastic scintillator when the beam hits a specific point located on top of the frontal scintillator area. We show results for configuration D in Fig.~\ref{conf}.} 
\end{figure}

\begin{figure}[!htb]
\centering
\includegraphics[width=21pc]{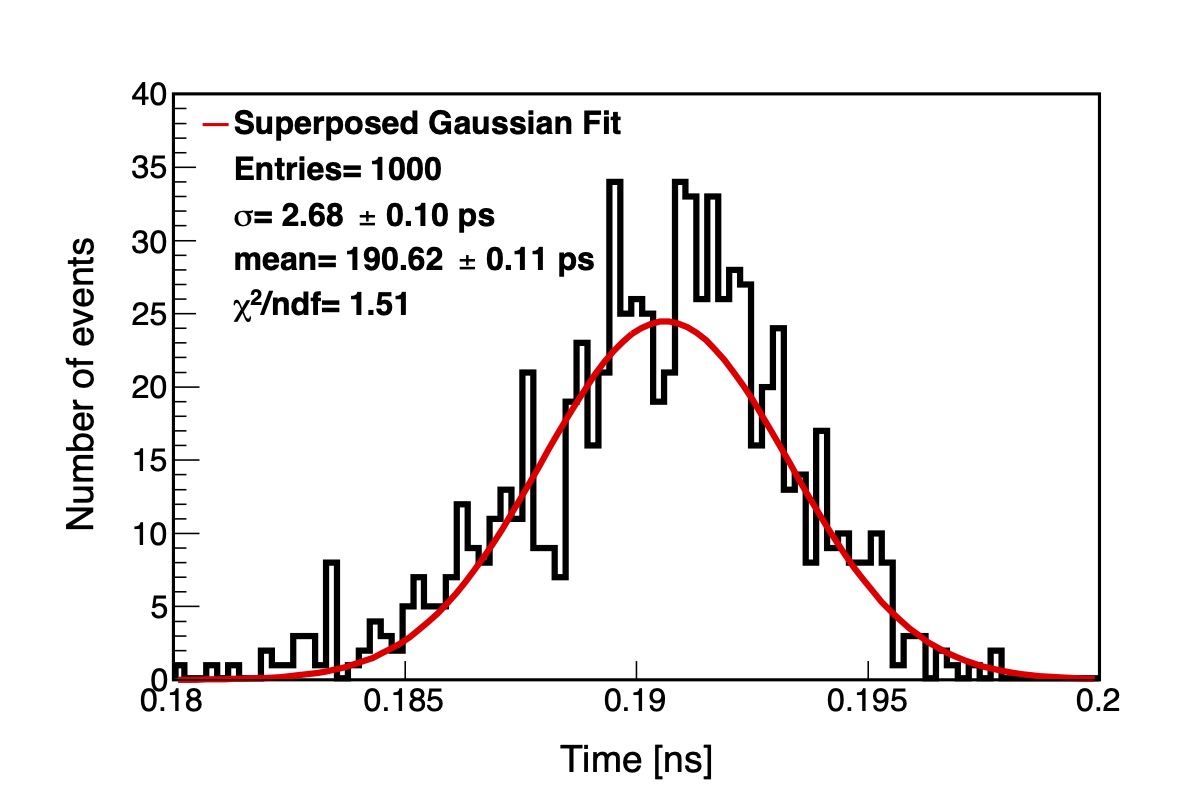}
\caption{\label{2individual}Time-of-flight distribution for the top interaction point, as in Fig. \ref{top}, but with only one scorer, which corresponds to configuration A in Fig.~\ref{conf}.}
\end{figure}

Figure~\ref{gaus} shows the distribution for the case of 4 scorers. The two peaks are due to the randomly distributed incidences 
all over the cell area; the same pattern is observed when working with the other configurations. To understand this effect, we performed two more simulations in which the beam hits one specific point of the cell; one where the particles hit the center of the cell and the other where the particles hit one corner of the cell. 
Figure~\ref{top} illustrates this scenario for the example of the time-of-flight distribution for the interaction in the center of the cell. Figure~\ref{2individual} shows the corresponding 
distribution for the case when only one scorer is simulated. The interaction at the corner leads to a time resolution of around 2.6~ps, but when the interaction is at the center of the cell this time increases to about 26~ps. This difference is due to the optical path that the photons travel from where they are produced until they reach the scorer. For the corner interaction, photons are reflected almost immediately and reach the scorer faster, while for the interaction at the center, the optical path for reflected photos is almost twice the distance to the edge.
We repeated this analysis for the other configurations which also led to approximately Gaussian peaks. Again, the  interval of the time resolution is equivalent for all cases, due to the significantly coarser resolution of the electronics. These results suggest that central interactions are inappropriate to obtain a lower time resolution.

We conclude that all four configurations and both materials are equivalent with an average value around 8~ps for interactions all over the frontal area.  Albeit this time is expected to be sensitive to the  location of the interaction point. We do not 
observe  appreciable differences between the time resolution obtained for each configuration. The difference is visible, however,
when considering the photon arrival time: for the case of one scorer this time is between 60--192~ps, decreasing to the interval 30--60~ps for the case of four scorers.
Therefore, we infer that the configuration with four scorers provides the best
intrinsic time resolution.

We also notice that if use was made of a SiPM with a larger effective area, for example one with a
$6 \times 6 ~{\rm mm}^2$ area, the intrinsic time resolution would remain essentially the same. Any possible improvement would be
of the order of a few ps. Hence our results for various arrangements of 
scorers would not show any significant improvement for the cell intrinsic time resolution.

\section{Implementation of the miniBeBe geometry in the MPDRoot Framework: Hits, Energy Deposit and Time-Of-Flight}\label{secVII}

\begin{figure}[!hbt]
\centering
\includegraphics[scale=0.15]{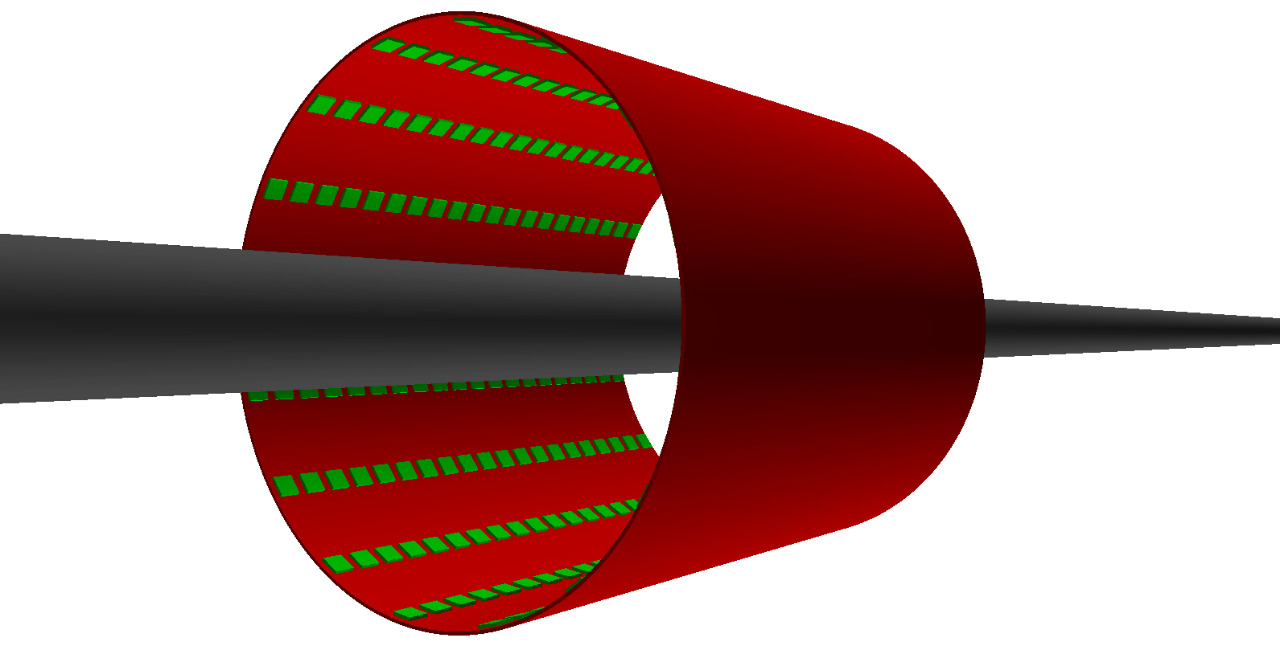}
\caption{Geometry of the miniBeBe as simulated within MPDRoot and rendered by the Event Display. Sixteen strips are arranged surrounding the interaction point of the MPD. Each strip consists of 20 squared plastic scintillators of size $20 \times 20 \times 3 ~\mathrm{mm}^3$, made of BC404. The simulated sensitive area has a
length of 60~cm and its diameter is 50 cm.}
\label{fig:mbb_strip}
\end{figure}

Using the official offline framework of the MPD, MPDRoot, we simulated the miniBeBe under the specifications described in Sec.~\ref{sec:secII}. Figure~\ref{fig:mbb_strip} shows the Geometry of the miniBeBe as simulated within MPDRoot, confirming that MPDRoot has the geometry implemented as per design. In order to test the implementation of the miniBeBe in the MPDRoot framework, we performed simulations of 950,000 Minimum Bias (MB) events (impact parameter $b = 0 - 15.98$ fm) for Bi + Bi collisions at $\sqrt{s_{NN}}=9$ GeV and 950,000 events for p + p collisions, using UrQMD~\cite{Urqmd1,Urqmd2}.

\begin{figure}[!hbt]
\centering
\includegraphics[scale=0.4]{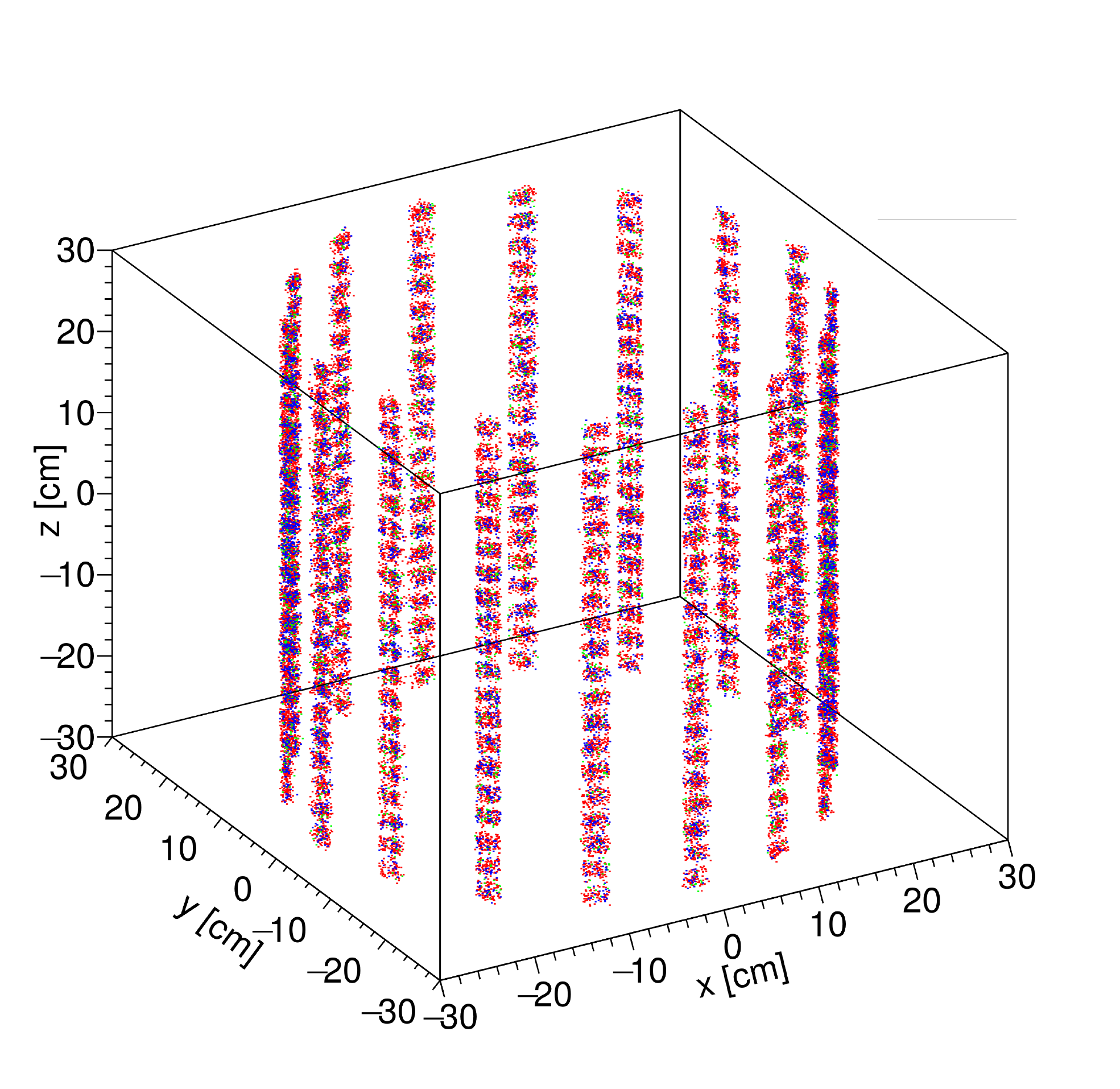}

\includegraphics[scale=0.2]{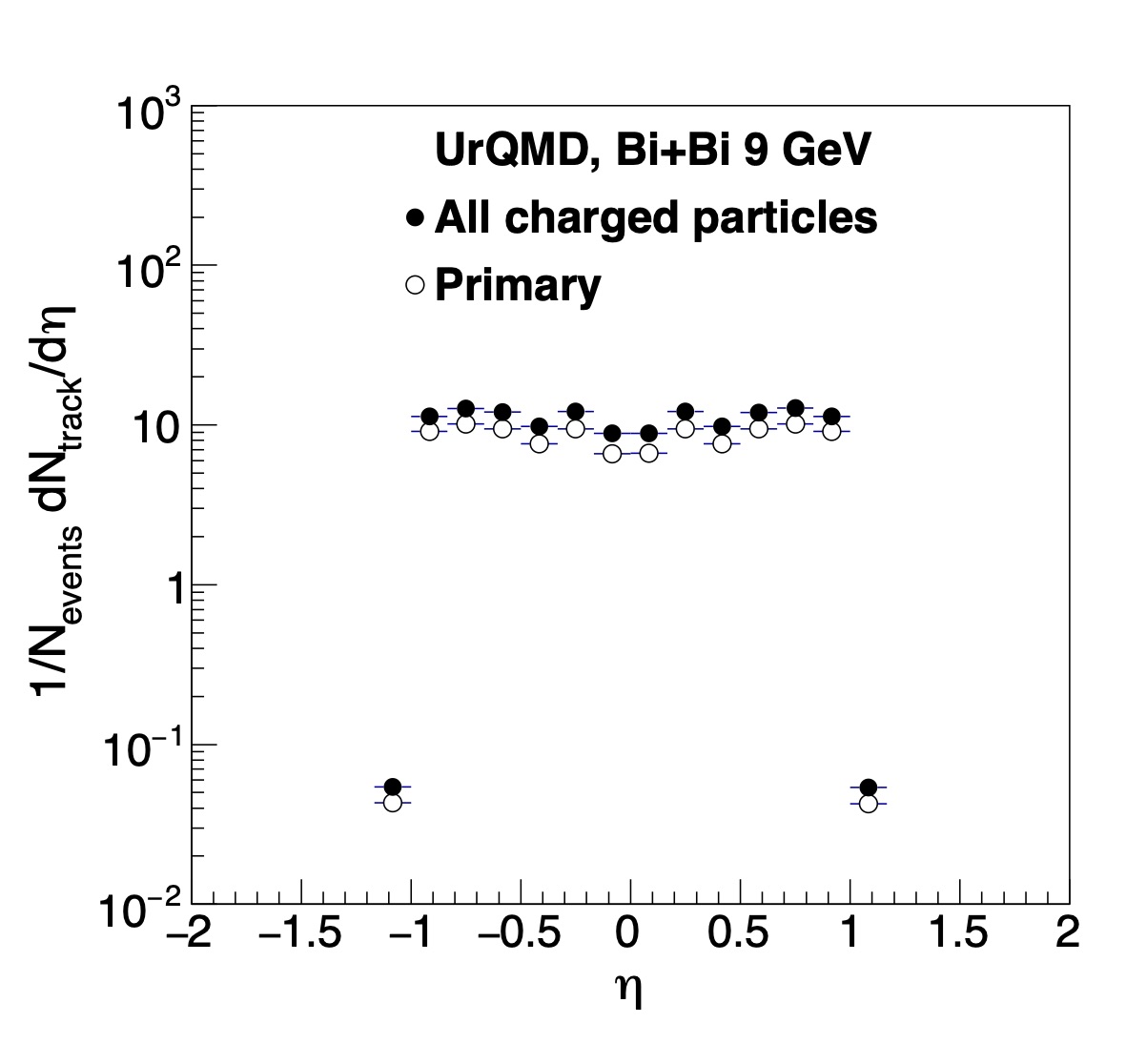}

\caption{Geometrical selection of the miniBeBe cells of tracks (MCTracks within MPDRoot) using 5000 events for Bi + Bi collisions at 9 GeV shown as hits in space (top), and the $\eta$ distribution of all charged particles and primary particles only (bottom), where we can verify that indeed the acceptance of the miniBeBe is $|\eta| < 1.1$.}
\label{fig:tracks1}
\end{figure}

\begin{figure}[!hbt]
\centering

\includegraphics[scale=0.17]{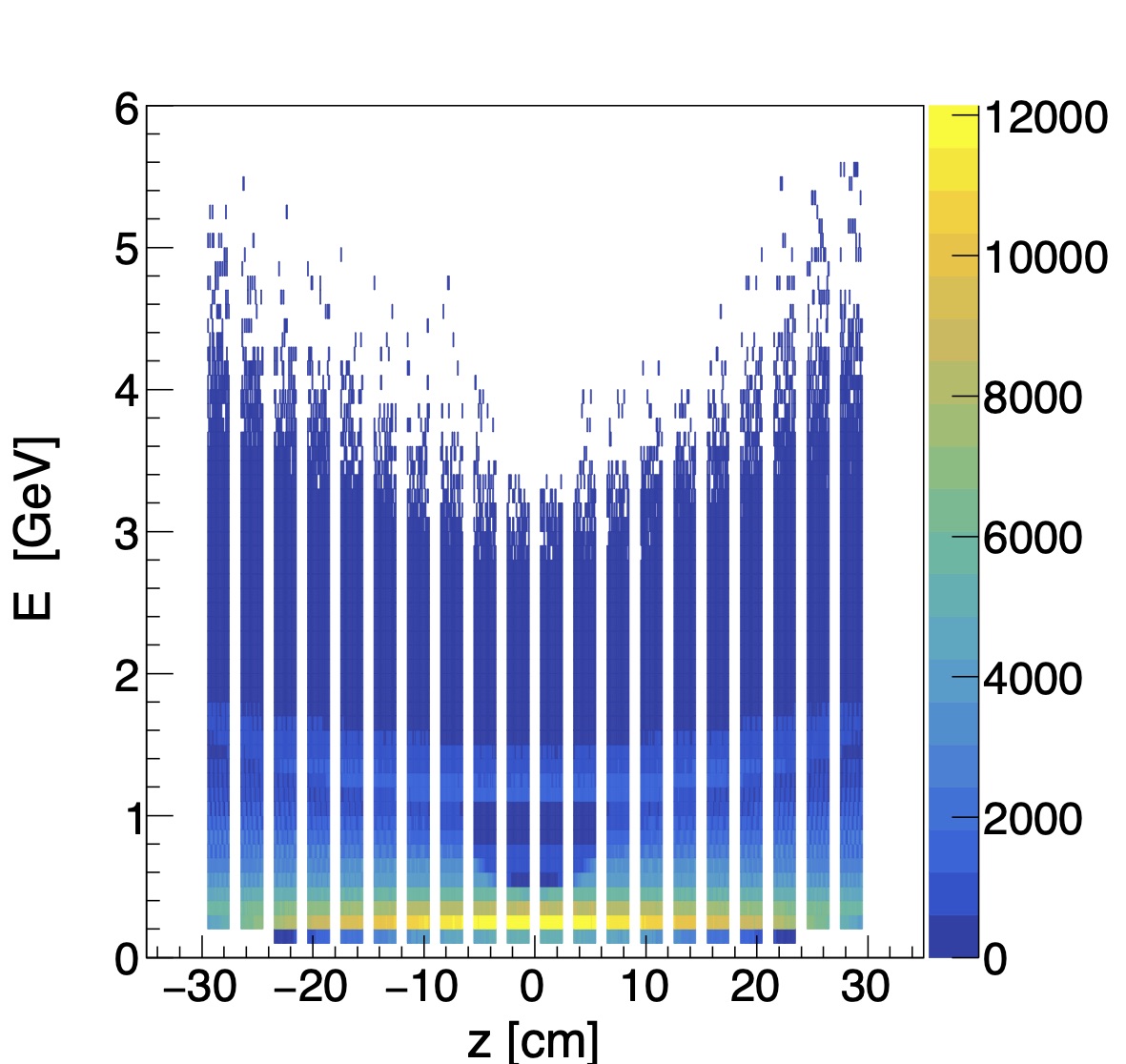}

\includegraphics[scale=0.17]{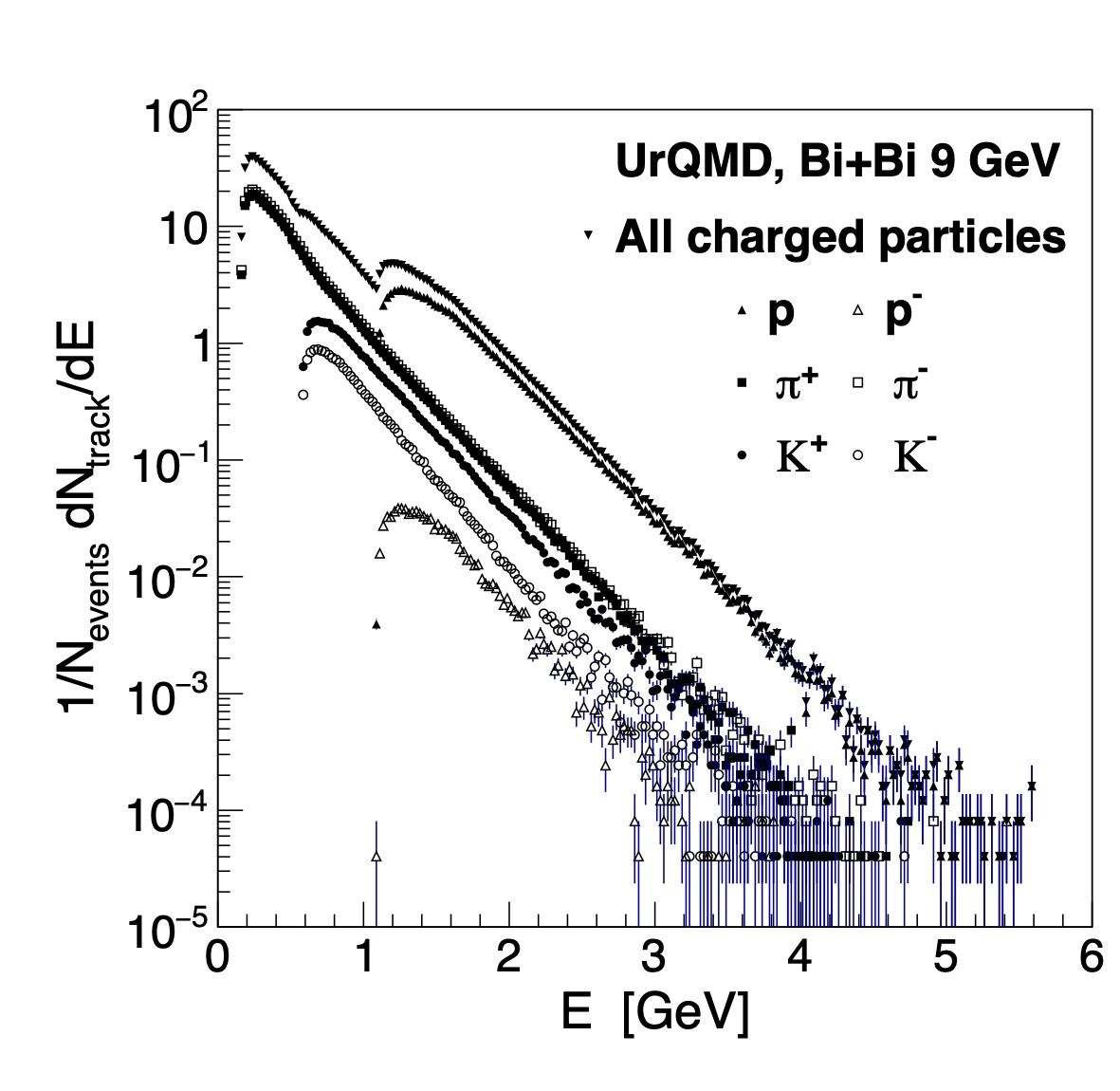}

\caption{Scatter plot distribution of particles with respect to the energy they carry at generation level within MCTracks when they reach the miniBeBe (top) and identified particle distributions (bottom), normalized by the number of events for a sample of Bi + Bi collisions at 9~GeV.}
\label{fig:tracks2}
\end{figure}

First we concentrate on the tracks selected in the geometrical acceptance of the miniBeBe and study the energy of particles hitting the detector cells, in order to compare with the energy deposited when we include the material. We perform a geometrical selection of the miniBeBe cells of tracks (MCTracks within MPDRoot) as shown for Bi + Bi collisions at 9 GeV in Fig.~\ref{fig:tracks1} with the hits in space (top) and with the $\eta$ distribution of all charged particles and primaries (bottom), where we can verify that indeed the acceptance of the miniBeBe occurs at $|\eta| < 1.1$. Then, in Fig.~\ref{fig:tracks2} we analyze the distribution of particles with respect to the energy of their tracks obtained from the MC and we show both the scatter plot (top) and the identified particle distributions (bottom). As expected, pions are most abundant in the lower energy domain of the spectra.

\begin{figure}[!hbt]
\centering
\includegraphics[scale=0.4]{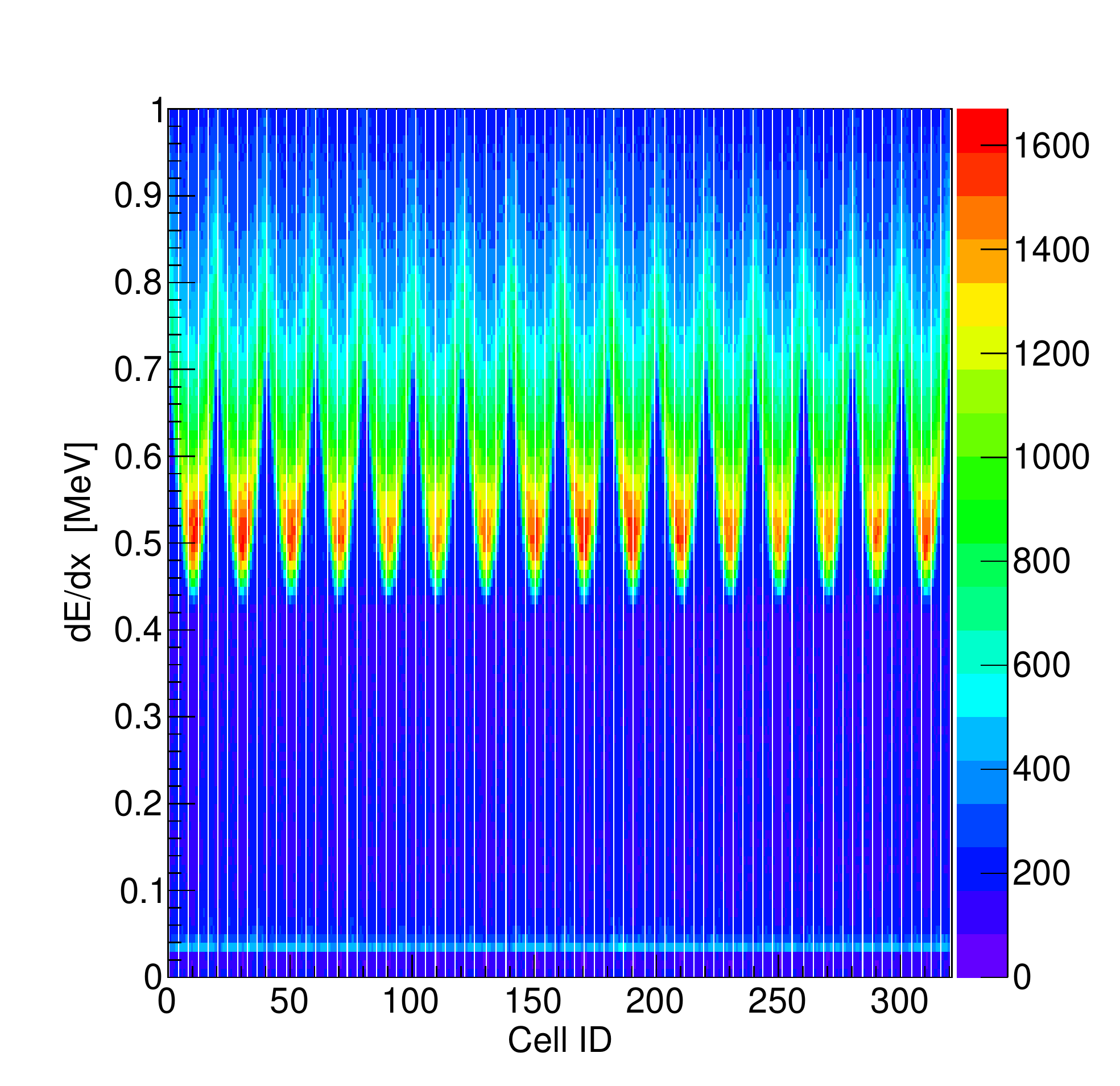}

\includegraphics[scale=0.4]{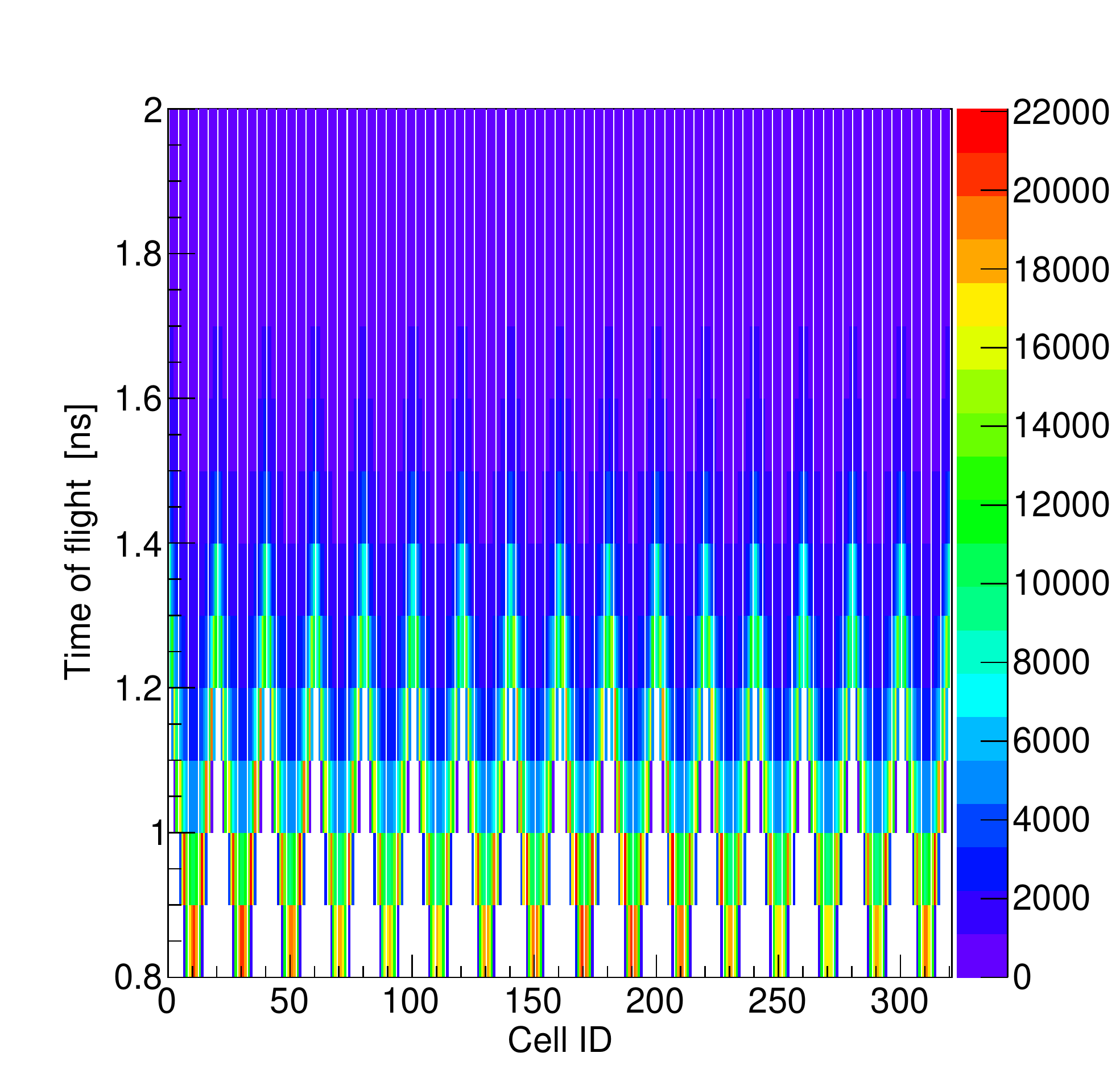}

\caption{Scatter plots for the hits in all the miniBeBe cells for the MB sample of Bi + Bi at 9 GeV. The upper panel shows the energy deposit and lower panel the time-of-flight for all hits. Given our convention to label the cells, the maximum of the energy deposit and the minimum time-of-flight happen for cells labeled by integer multiples of 10. }
\label{fig:mbb_scatter}
\end{figure}


Next, we perform a hit level analysis of the energy deposit and of the time-of-flight using this geometry. Figure~\ref{fig:mbb_scatter} shows scatter plots for the miniBeBe where we indicate the cell identification on the horizontal axis. Notice that there is a band regularity corresponding to the cells per strip that is reflected in the next part of this analysis.  Notice also that if we compare the energy scale of charged particles given in Fig.~\ref{fig:tracks2} and the scale for energy deposit in the miniBeBe in Fig.~\ref{fig:mbb_scatter}, we can see that most charged particles deposit by far less than 1\% of their energy in the miniBeBe. The scatter plots serve as a test of the coverage of the cells in a strip and shows the uniformity of the coverage.



 In order to optimize the coverage of the cells for the miniBeBe, we monitor the strip-averages of the number of hits, the energy deposit and the time-of-flight per cell along a miniBeBe strip, by averaging over the sixteen strips in this nominal configuration. Since each strip has 20 cells, we use the notation for evenly-spaced cells 1 to 20 to refer to their location from $z=-30$~cm to $z=+30$~cm. These studies enabled an optimized trigger design and improvements, as will be summarized at the end of this section and as will be reported in detail in the next sections.
 
 We now report the strip-average number of hits, energy deposit and time-of-flight in miniBeBe for Bi + Bi collisions at $\sqrt{s_{NN}}=9$ GeV for a MB sample and for samples in three centrality classes: 0-20\%, 40-60\% and 80-100\% (where \lq\lq \%\rq\rq ~represents the percentage of the total cross-section), which correspond to impact parameter ranges $b=0-4.80$ fm, $b=7.36-9.97$ fm and $b=12.97-15.98~{\rm fm}$, respectively. We also report these strip-average quantities for p + p collisions at $\sqrt{s_{NN}}=4, 9, 11$ GeV.

\begin{figure}[hbt!]
\centering

\includegraphics[trim=0.2cm 0.3cm 0.3cm 0.5cm ,clip,scale=0.2]{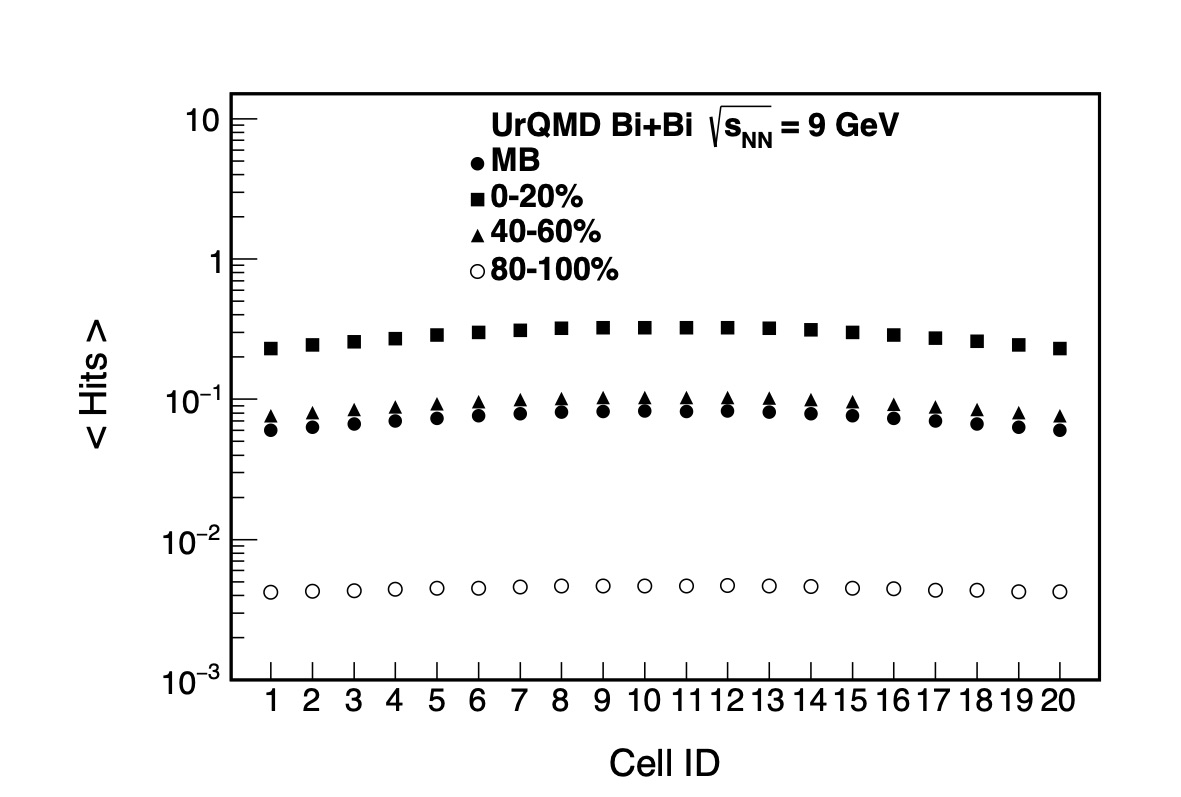}

\includegraphics[trim=0.2cm 0.3cm 0.3cm 0.5cm ,clip,scale=0.2]{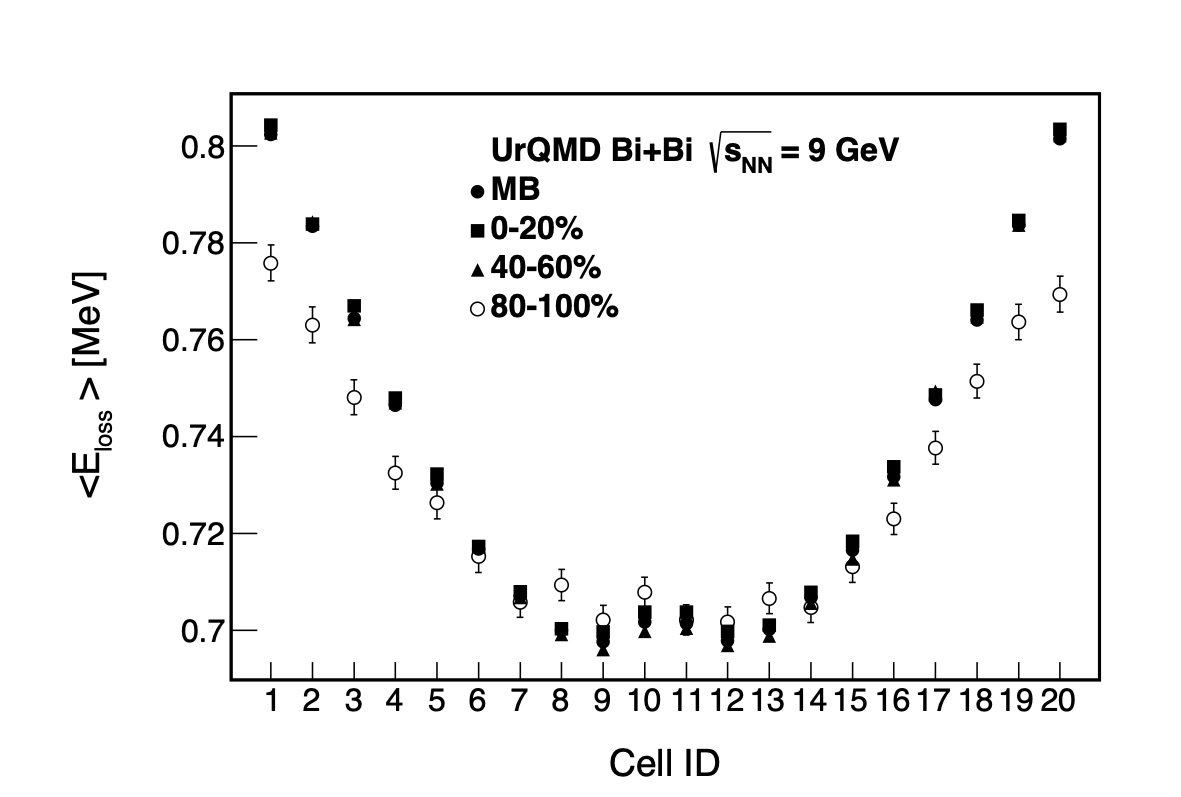}

\includegraphics[trim=0.2cm 0.3cm 0.3cm 0.5cm ,clip,scale=0.2]{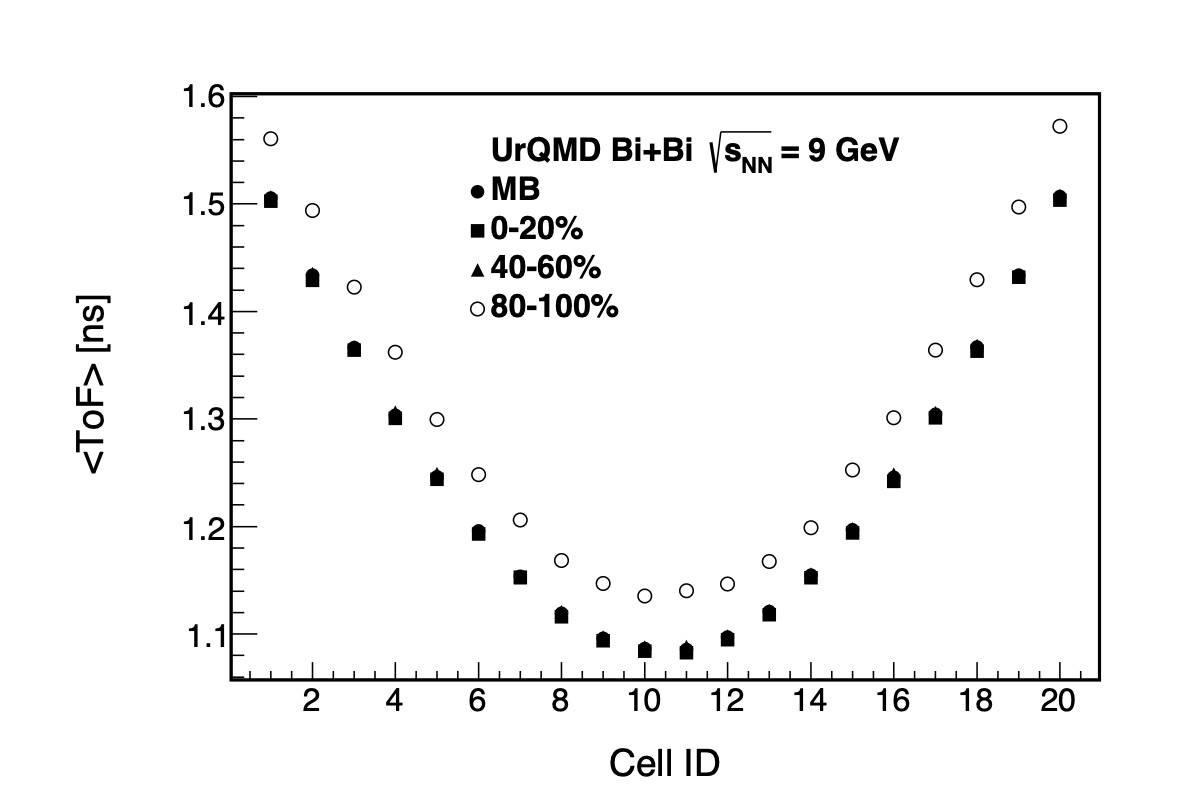}

\caption{Strip-average of the number of hits (top), energy deposit (center) and time-of-flight (bottom) per cell for the miniBeBe in Bi + Bi collisions at 9 GeV. We show results for the MB samples ($b = 0 - 15.8$~fm), as well as for three different centrality classes.}
\label{fig:mbb_avgs}
\end{figure}

Figure~\ref{fig:mbb_avgs} shows the strip-average number of hits (top), energy deposit (center) and time-of-flight (bottom) per cell along a miniBeBe strip for Bi + Bi at 9 GeV. All the panels include the MB and the centrality classes results. We notice that we have on average almost 3 hits per strip in the most central collisions, down to 1 hit per strip in the semi-central collisions and less than 1 hit on average for the peripheral collisions. Considering that the miniBeBe has 16 strips, we expect the highest miniBeBe efficiency at around 48 hits per event for Bi + Bi central collisions. In the center panel, we note that we have an average energy deposited per miniBeBe cell of at most $\simeq 0.8$~MeV for all centrality classes. So we expect the miniBeBe to withstand, on average, 16 MeV of energy deposited per strip. At the bottom, the panel for the average time-of-flight shows that for the central miniBeBe cells (around $z=0$) we have an average below 1.3 ns. Note also that we can reach time-of-flight averages of (slightly) less than 1.1 ns. This sets the benchmark analysis for the trigger capabilities of the miniBeBe in the next section, where we compare leading time vs.\ average time results for both Bi + Bi and p + p collisions.

\begin{figure}[hbt!]
\centering

\includegraphics[trim=0.2cm 0.3cm 0.3cm 0.5cm ,clip,scale=0.2]{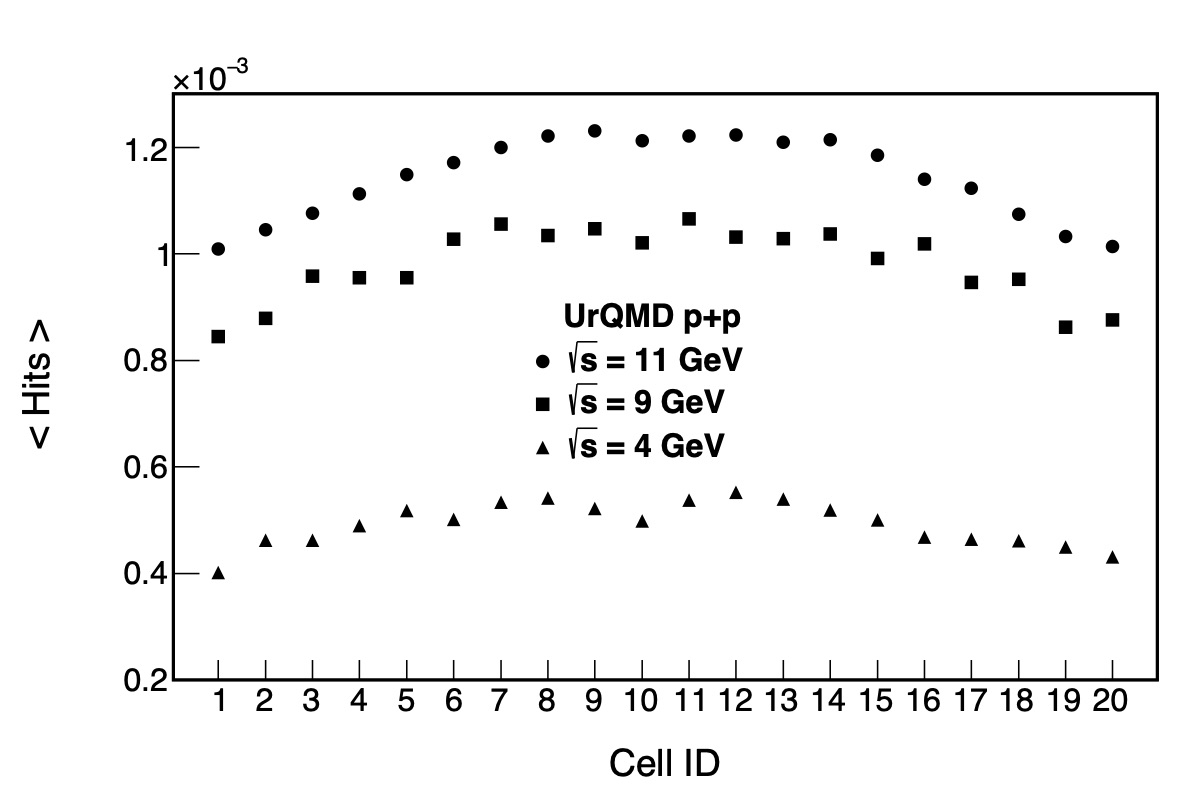}
\includegraphics[trim=0.2cm 0.3cm 0.3cm 0.5cm ,clip,scale=0.2]{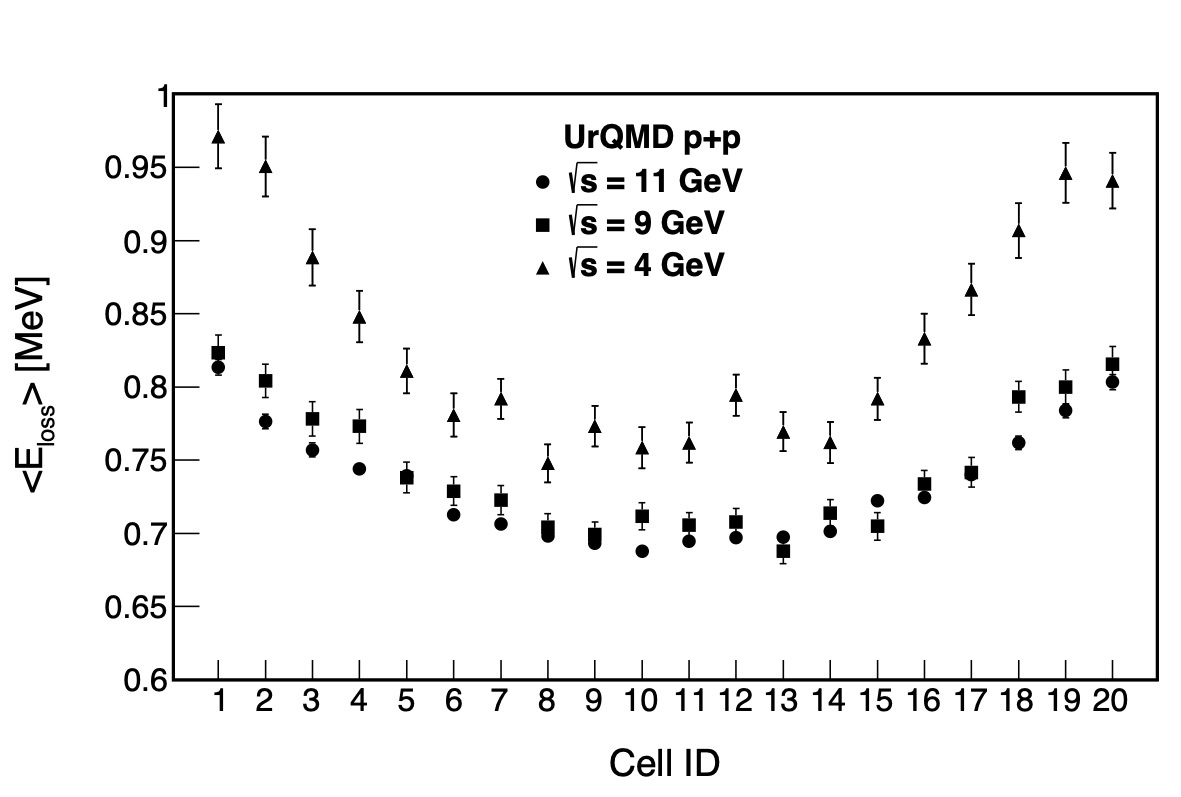}
\includegraphics[trim=0.2cm 0.3cm 0.3cm 0.5cm ,clip,scale=0.2]{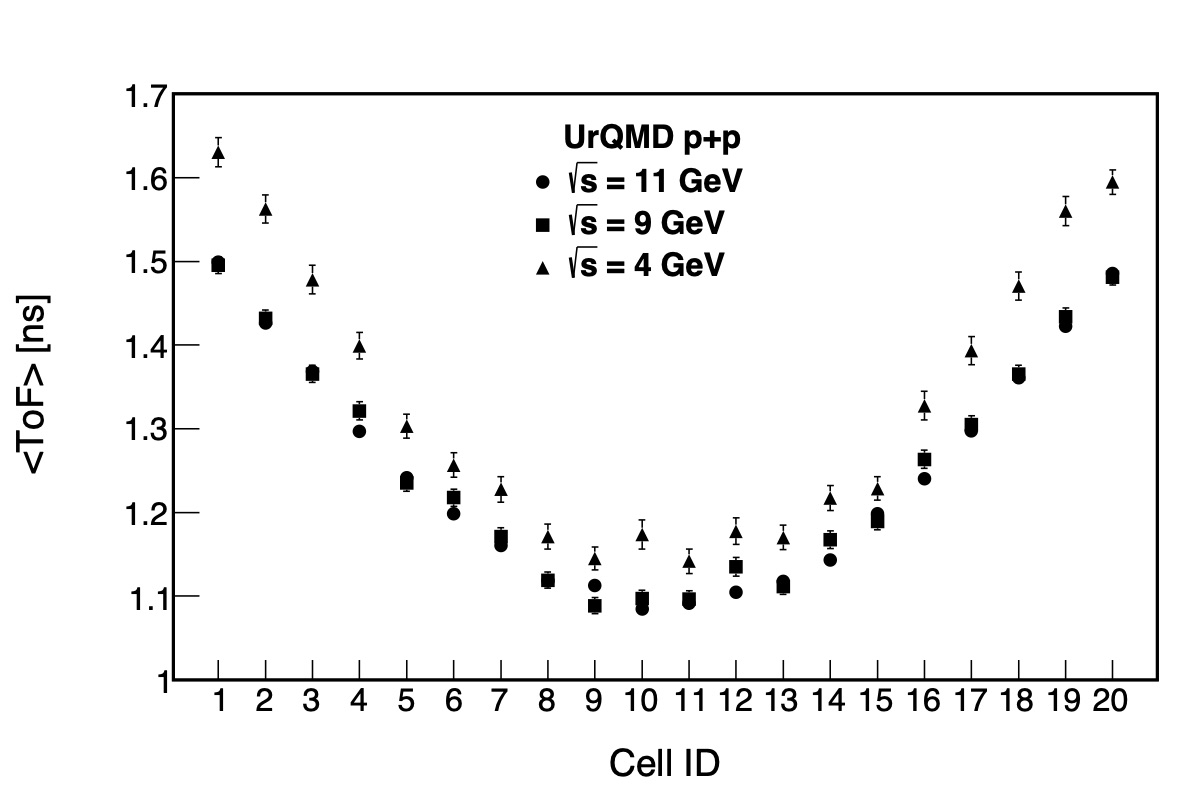}

\caption{Strip-average of the number of hits (top), energy deposit (middle) and time-of-flight (bottom) per cell for the miniBeBe in p + p collisions at 4, 9 or 11 GeV.}
\label{fig:mbb_avgpp}
\end{figure}

Notice also that peripheral heavy ion collisions should be comparable to p + p collisions. For completeness, Fig.~\ref{fig:mbb_avgpp} shows the average number of hits, energy loss and time-of-flight using 950,000 p + p collision events at $\sqrt{s} = 4, 9, 11$ GeV that we generated using UrQMD and we transported through miniBeBe using MPDRoot. We notice that even though the average number of charged particles in p + p is well below that of Bi + Bi collisions, they deposit more energy in the detector. Overall, we have a similar scale of energy deposit per cell in p + p and in Bi + Bi collisions, so our findings are summarized as follows: for Bi + Bi at $\sqrt{s_{NN}}=$ 9 GeV, we have shown that the average number of hits, the average energy deposit and the average time-of-flight per design geometry of the miniBeBe, happens within an average time-of-flight between 1.1 and 1.6 ns. Moreover the length of the detector covers the region with the highest average hits per event with no more than 16 MeV of energy deposit per strip. We have also verified, using the energy deposit of charged particles, that the miniBeBe has a small occupancy, of order 2\%, .


\begin{figure}[!htb]
\centering

\includegraphics[trim=0.3cm 0.3cm 0.3cm 0.5cm ,clip,scale=0.2]{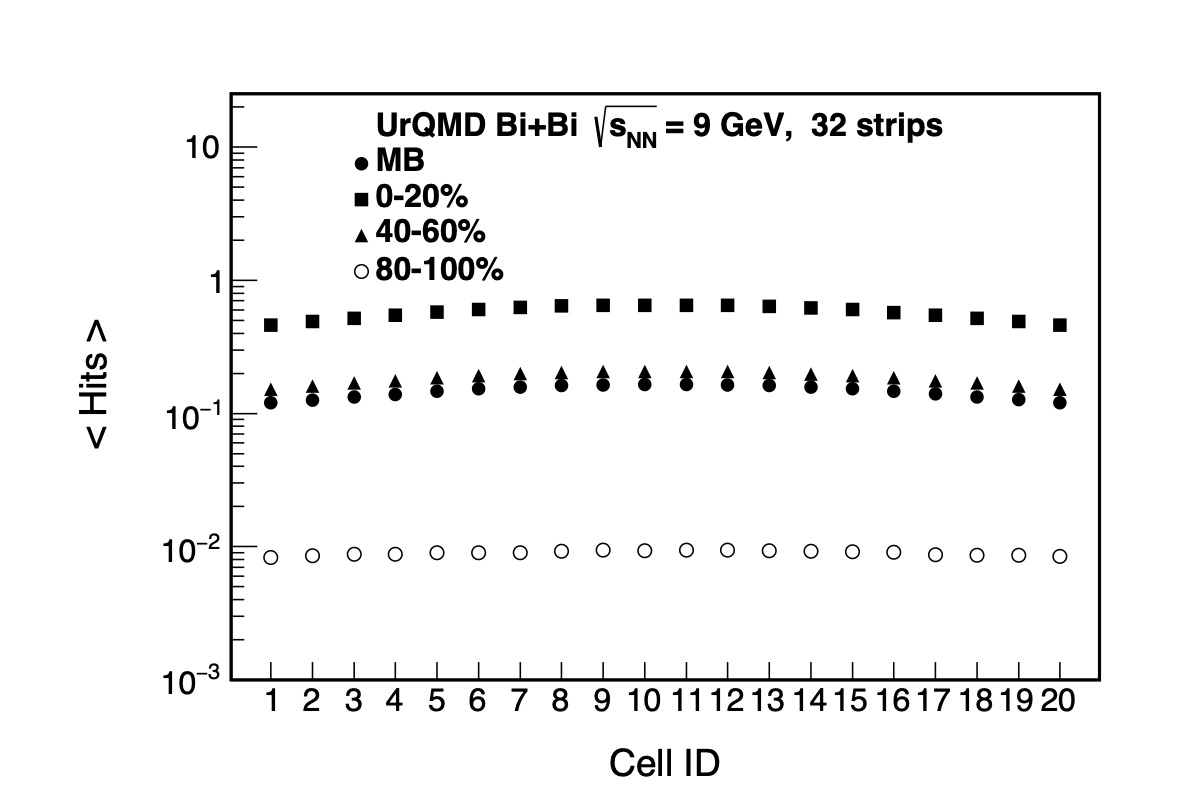}
\includegraphics[trim=0.3cm 0.3cm 0.3cm 0.5cm ,clip,scale=0.2]{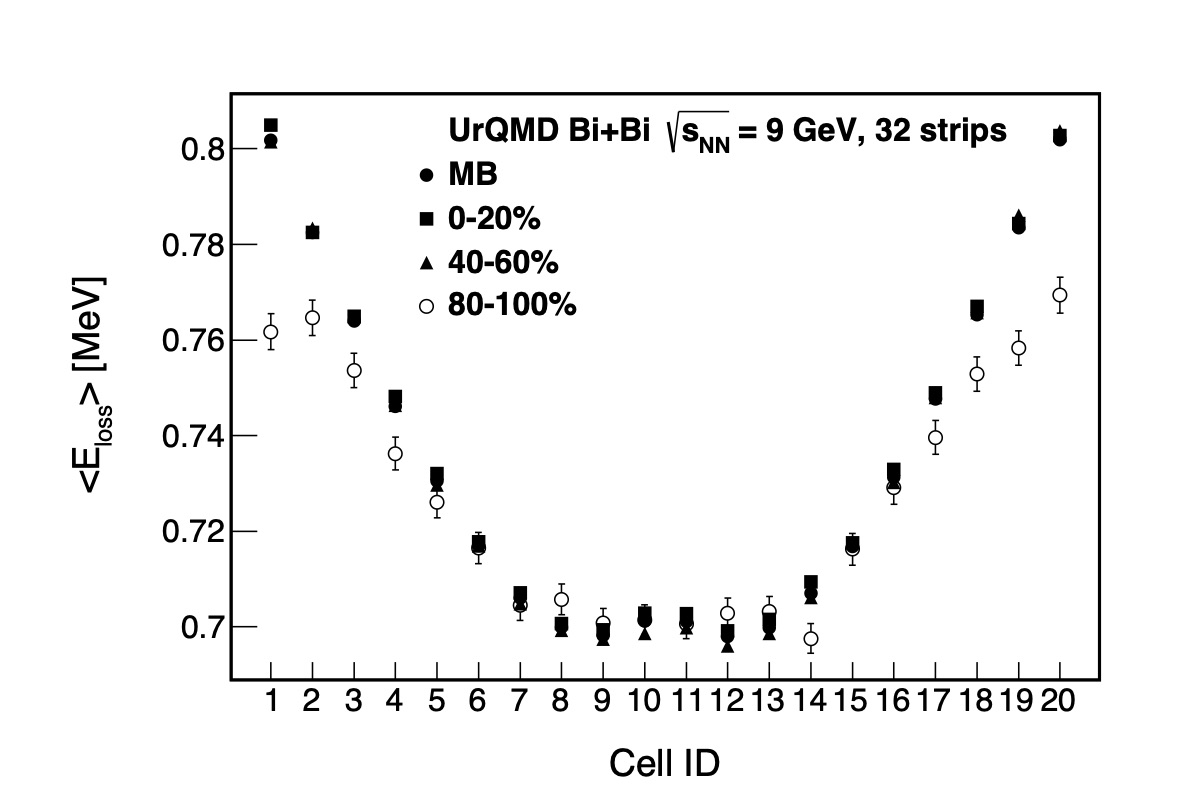}
\includegraphics[trim=0.3cm 0.3cm 0.3cm 0.5cm ,clip,scale=0.2]{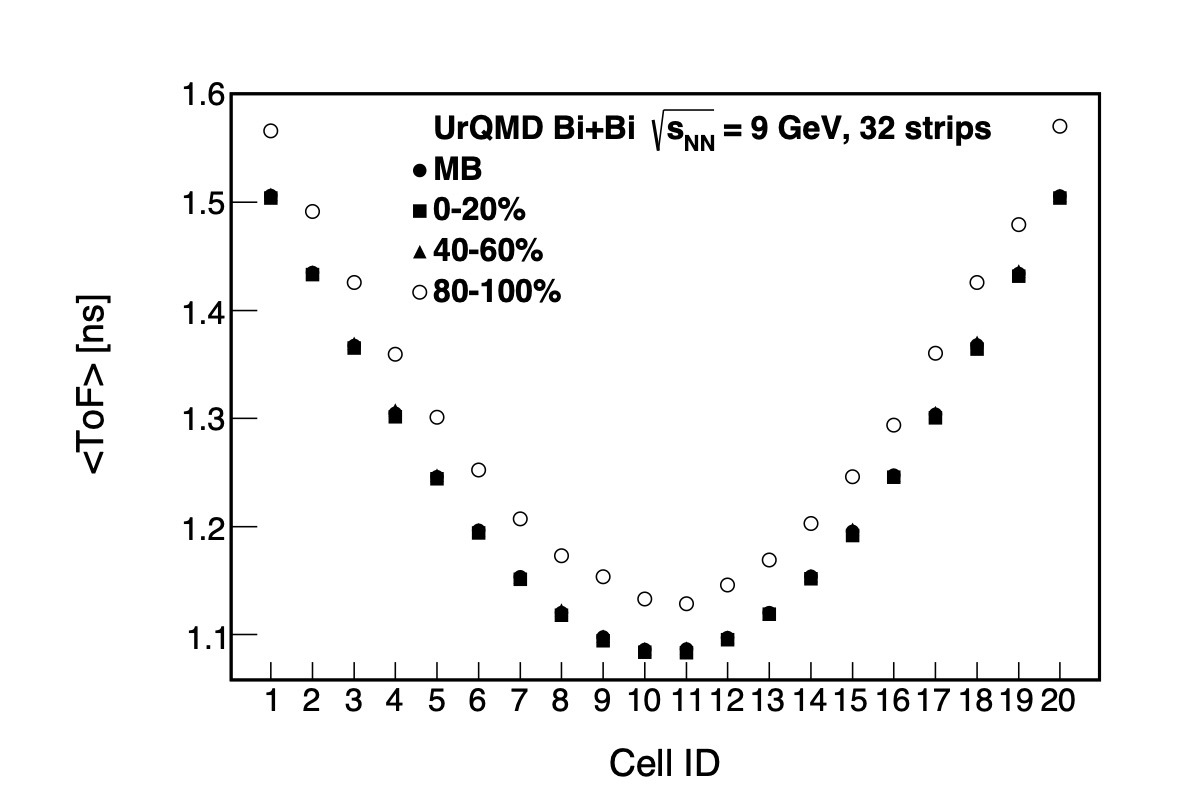}

\caption{Strip-average for the upgraded geometry with 32 strips, of the number of hits (top), energy deposit (middle) and time-of-flight (bottom) per cell for the miniBeBe in Bi + Bi collisions at $\sqrt{s_{NN}}=$ 9 GeV.}
\label{fig:mbb_upgrade-AA}
\end{figure}

\begin{figure}[!htb]
\centering

\includegraphics[trim=0.3cm 0.3cm 0.3cm 0.5cm ,clip,scale=0.2]{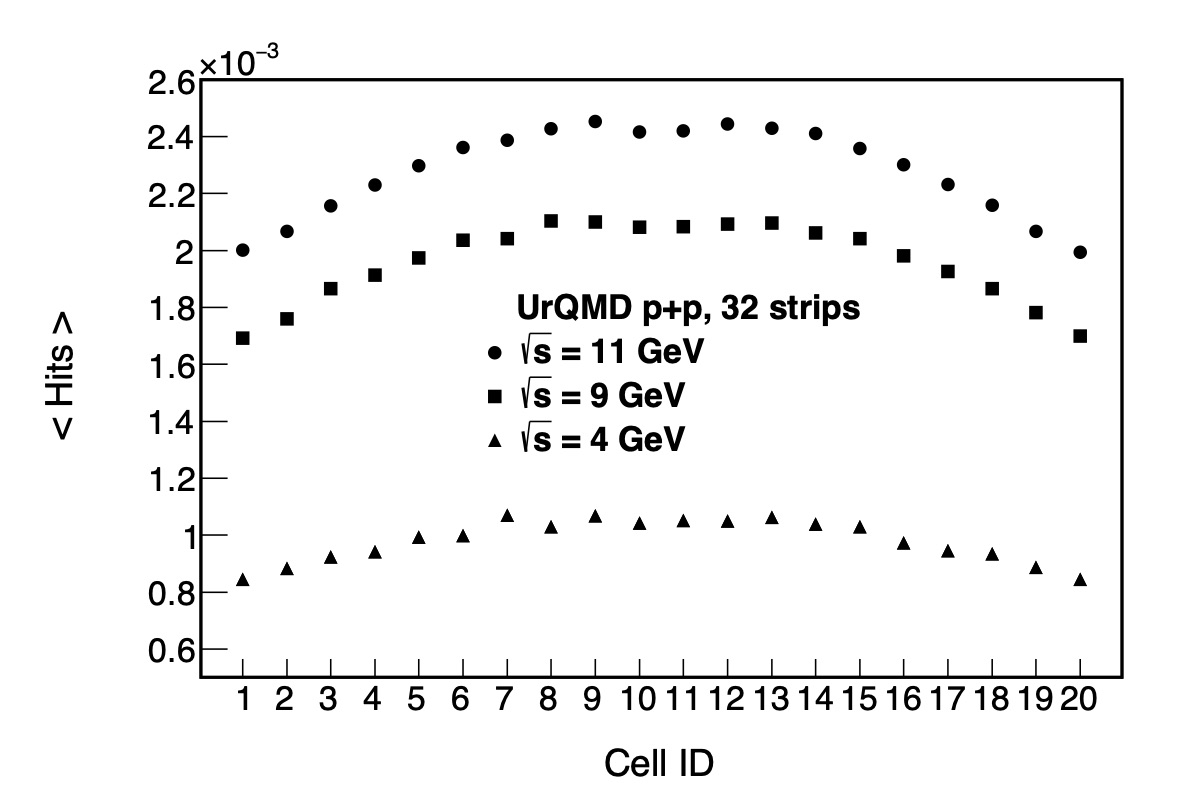}
\includegraphics[trim=0.3cm 0.3cm 0.3cm 0.5cm ,clip,scale=0.2]{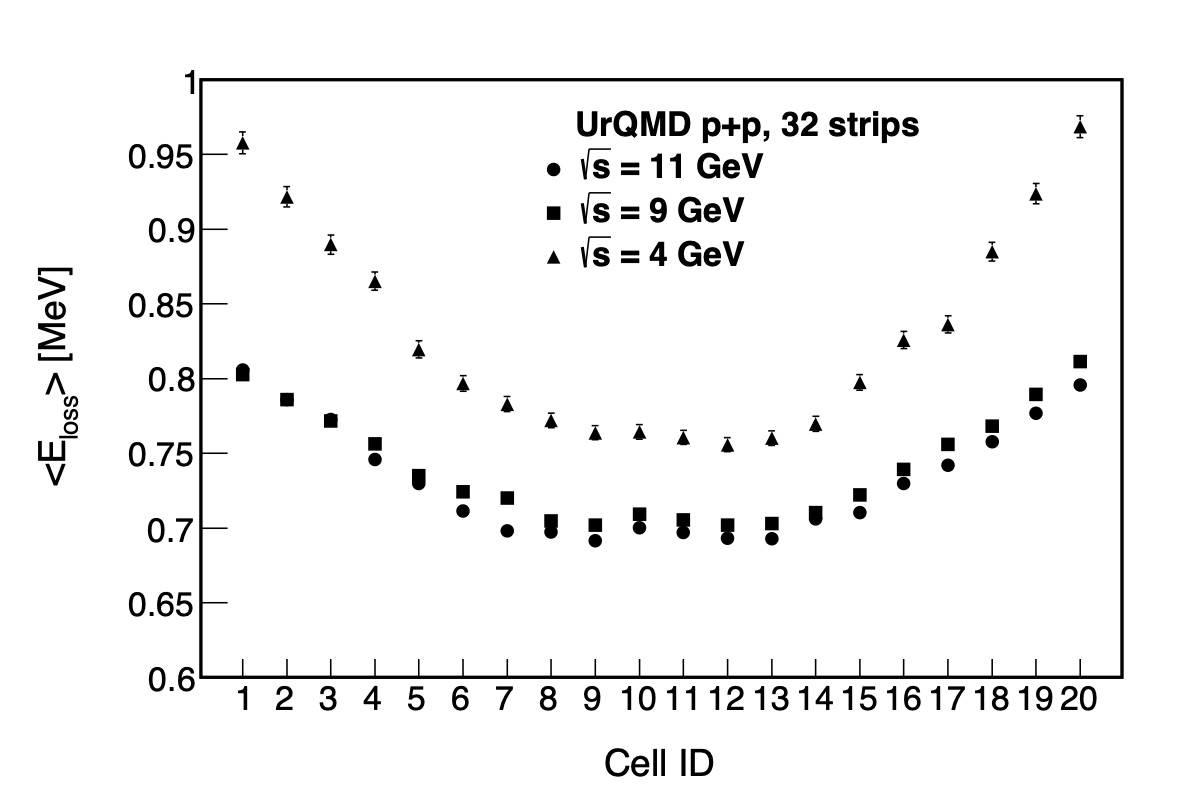}
\includegraphics[trim=0.3cm 0.3cm 0.3cm 0.5cm ,clip,scale=0.2]{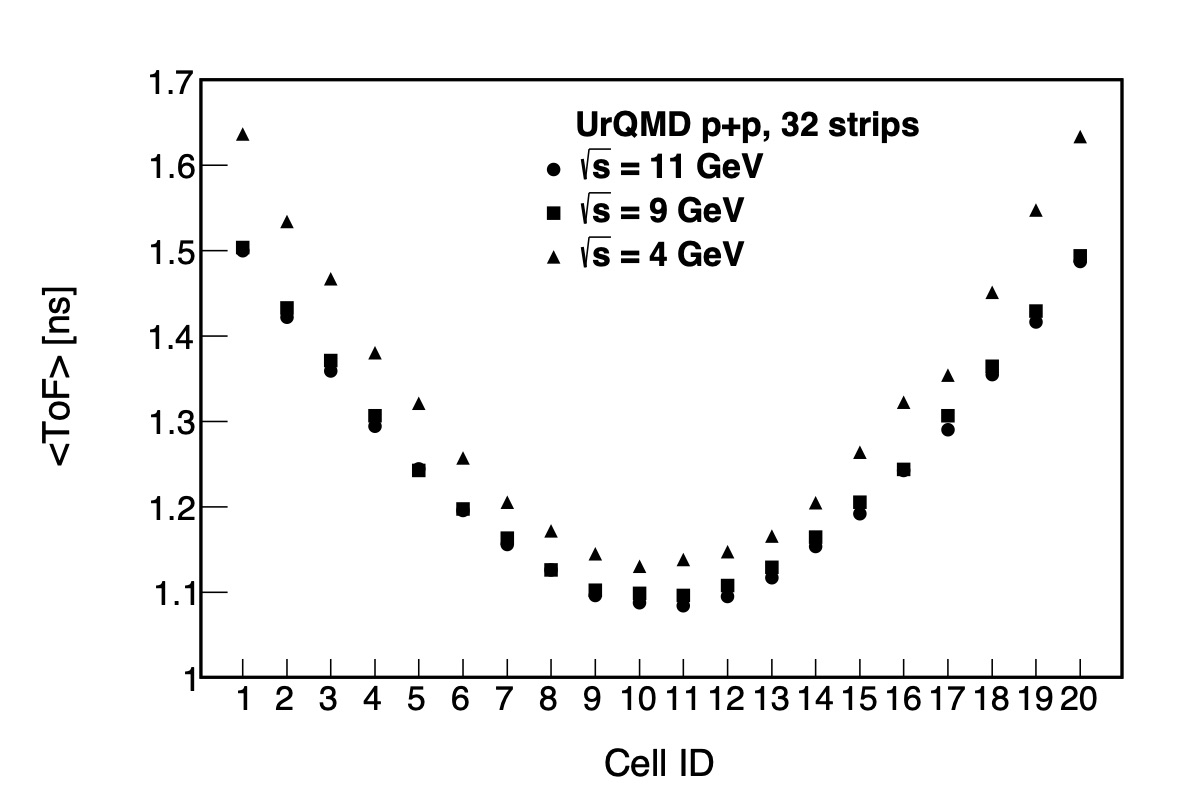}

\caption{Strip-average for the upgraded geometry with 32 strips, of the number of hits (top), energy deposit (middle) and time-of-flight (bottom) per cell for the miniBeBe in p + p collisions at 4, 9 and 11 GeV.}
\label{fig:mbb_upgrade-pp}
\end{figure}

To conclude this section, we comment on possible and immediate improvements for the miniBeBe design, that still conform to current space  availability in MPD, but that are contingent upon further financial support. In Figs.~\ref{fig:mbb_upgrade-AA} and ~\ref{fig:mbb_upgrade-pp} we show the expected increase of the average number of hits in the miniBeBe when doubling the number of strips. We use $5 \times 10^5$ events for Bi + Bi collisions at  $\sqrt{s_{NN}}=$ 9 GeV generated with UrQMD and for p + p at $\sqrt{s_{NN}}=$ 4, 9, 11 GeV, transported with MPDRoot through an upgraded miniBeBe that now has 32 strips.

The impact on the baseline design of miniBeBe is summarized in Table~\ref{tab:avghitscomp}. We emphasize the summary for the average number of hits per cell and for the complete miniBeBe detector, in both the 16-strip and 32-strip designs. As expected, the average number of total hits in miniBeBe doubles when proceeding from the 16-strip to the 32-strip design. Since each strip has 20 cells, the complete detector average hit range is obtained with a factor of $20\times 16$ and $20 \times 32$, for each geometry, respectively.

\begin{table}[!h]
    \centering
    \begin{tabular}{|c|c|c|c|c|}
    \hline
 UrQMD     & $\langle \mbox{Hits} \rangle$ &  strips & 0-20\%  & 80-100\%  \\
    \hline
            & per  & 16   & 0.2294 - 0.3248 & 0.0042 - 0.0047  \\
\cline{3-5}
 Bi + Bi  & cell & 32    & 0.2294 - 0.3250 & 0.0041 - 0.0047  \\
\cline{2-5}
\cline{2-5}
       9 GeV &  complete & 16    & 73.40 - 103.94 &  1.34 - 1.50 \\
       \cline{3-5}
         &  detector & 32    & 146.81 - 208.03 & 2.65 - 3.01 \\
\hline\hline
        UrQMD    & $\langle \mbox{Hits} \rangle$  & strips & 4 GeV & 11 GeV \\
    \hline
             & per   & 16    & 0.00043 - 0.00055 & 0.00100 - 0.00122 \\
             \cline{3-5}
    p + p     & cell & 32     & 0.00042 - 0.00058 & 0.00099 - 0.00122 \\
\cline{2-5}
\cline{2-5}
         &  complete  & 16     & 0.138 - 0.176 & 0.320 - 0.390 \\
         \cline{3-5}
        &  detector  & 32    & 0.269 - 0.339  & 0.637 - 0.784  \\
\hline
    \end{tabular}
    \caption{Overview of average number of hits in the miniBeBe as shown in Figs.~\ref{fig:mbb_avgs}, ~\ref{fig:mbb_avgpp}, ~\ref{fig:mbb_upgrade-AA} and ~\ref{fig:mbb_upgrade-pp}. For Bi + Bi at $\sqrt{s_{NN}}=$ 9~GeV, and p + p at $\sqrt{s_{NN}}=$ 4 and 11~GeV, we show the range of average number of hits per cell and of the complete detector. We show both the 16 and 32 miniBeBe geometry results and note that, as expected, the latter doubles the average number of total hits in the detector}. Since each strip has 20 cells, the complete detector average hit range is obtained with a factor of $20\times 16$ and $20 \times 32$, for each geometry.
    \label{tab:avghitscomp}
\end{table}

\section{Simulations for the MiniBeBe: Trigger Capabilities}\label{secVIII}

We used UrQMD \cite{Urqmd1,Urqmd2} for Bi + Bi collisions and beam-gas interactions. For Bi + Bi collisions, a sample of 9,000 MB events with a centrality range between 0 and 90\% was generated. For beam-gas interactions we simulated p+O collisions at $\sqrt{s_{NN}}=9$ GeV with a vertex position at $\pm 19$ m along the $z-$axis and a width of $\pm 3.5$ m. For these purposes, we considered the particle's velocity to be between $0.7c$ and $c$. 

\begin{figure}[!hbt]
\centering
\includegraphics[scale=0.2]{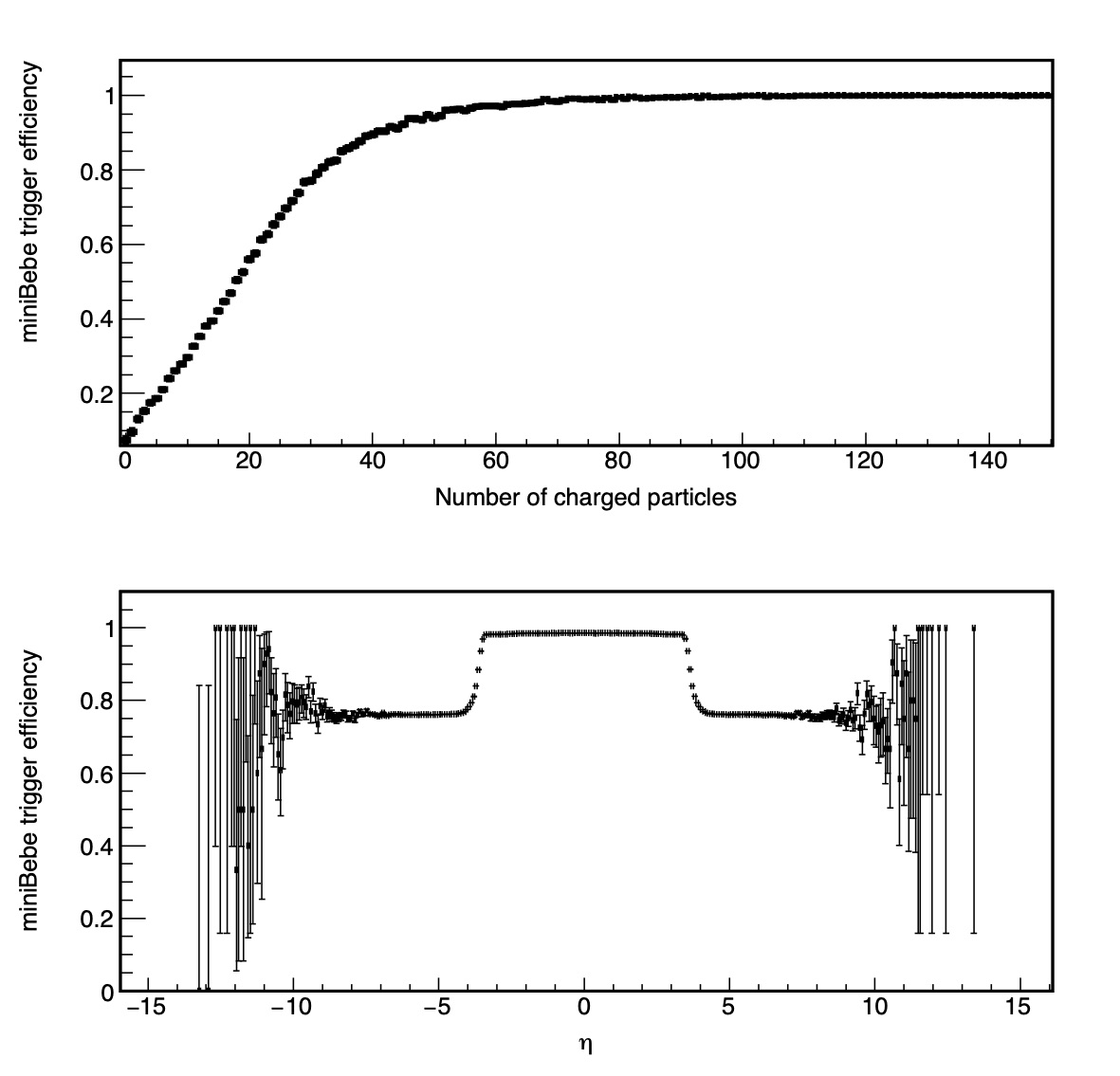}

\caption{MiniBeBe trigger efficiency as a function of the charged particle multiplicity (top) and pseudo-rapidity (bottom).}
\label{fig:minibebe_efficiency}
\end{figure}

The simulation was done to evaluate the trigger capabilities of the miniBeBe for heavy ion collisions, and to be used as a beam-gas interactions veto. Trigger efficiencies for miniBeBe have been obtained for Bi + Bi collisions at $\sqrt{s_{NN}}=9$ GeV. Figure \ref{fig:minibebe_efficiency} shows the trigger efficiency considering that at least one charged pion hits the miniBeBe. For low charged particle multiplicity events ($\lesssim 25$ charged particles), the miniBeBe trigger efficiency is less than 60\%. This behavior is due to the forward events that UrQMD generates, with few charged pions produced in the central barrel region. If we consider only events with charged particles within the miniBeBe detector acceptance ($|\eta|<1.01$), the trigger efficiency increases up to $\simeq 100$\%. In this case, the miniBeBe trigger efficiency is expected to be above 90\% for events with at least 50 charged particles, see Fig.~\ref{fig:minibebe_efficiency}.

\begin{figure}[!htb]
\centering
\includegraphics[scale=0.15]{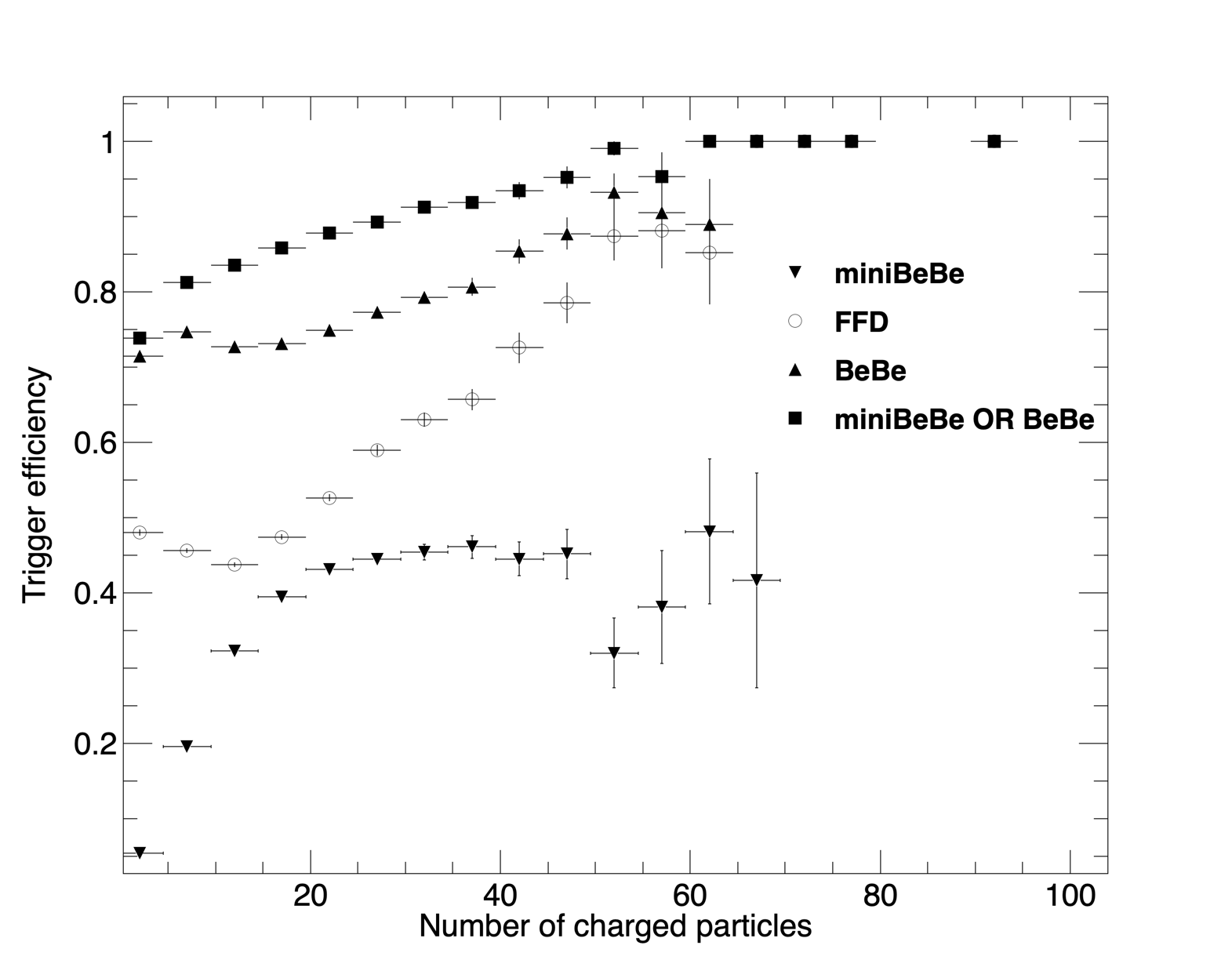}

\caption{Trigger efficiency as a function of the charged particle multiplicity for miniBeBe, BeBe and FFD detectors. Simulations are performed using UrQMD for p + p collisions with $\sqrt{s_{NN}}=9$ GeV.}
\label{fig:trigger_efficiency}
\end{figure}

Furthermore, we can compare the miniBeBe trigger efficiency with the one expected from the FFD~\cite{FFDpaper} and with the one expected from the proposed BeBe detector~\cite{BeBe}.  This comparison is shown in Fig.~\ref{fig:trigger_efficiency} for p + p collisions at $\sqrt{s_{NN}}=9$ GeV. Notice that the FFD trigger efficiency is higher than that of the miniBeBe, but it becomes smaller than 50\% for p + p events with less than 20 charged particles. For such events, the inclusion of the proposed BeBe detector in the MPD array increases these trigger capabilities. Moreover, combining the information of the miniBeBe and the BeBe detectors, the trigger efficiency for p + p low multiplicity events is at least 80\% for multiplicities where the FFD is not efficient.

\subsection{Multiplicity}\label{secVII_multiplicity}

At this stage of development of the miniBeBe in the MPDRoot frame, we extract the information of the physical interaction of particles produced in heavy-ion collisions at NICA energies using the volume of the miniBeBe which is sensitive to hits. Hits in the miniBeBe are produced when a Monte Carlo track enters into the active sensitive volume, without any restriction on the deposited energy. This is the standard definition of a hit in MPDRoot. The simplest information that we can extract from miniBeBe simulations is the number of hits per event and its corresponding time information. In this case, we assume that the number of hits in the miniBeBe can be taken as a raw multiplicity.

\begin{figure}[!hbt]
\centering
\includegraphics[scale=0.2]{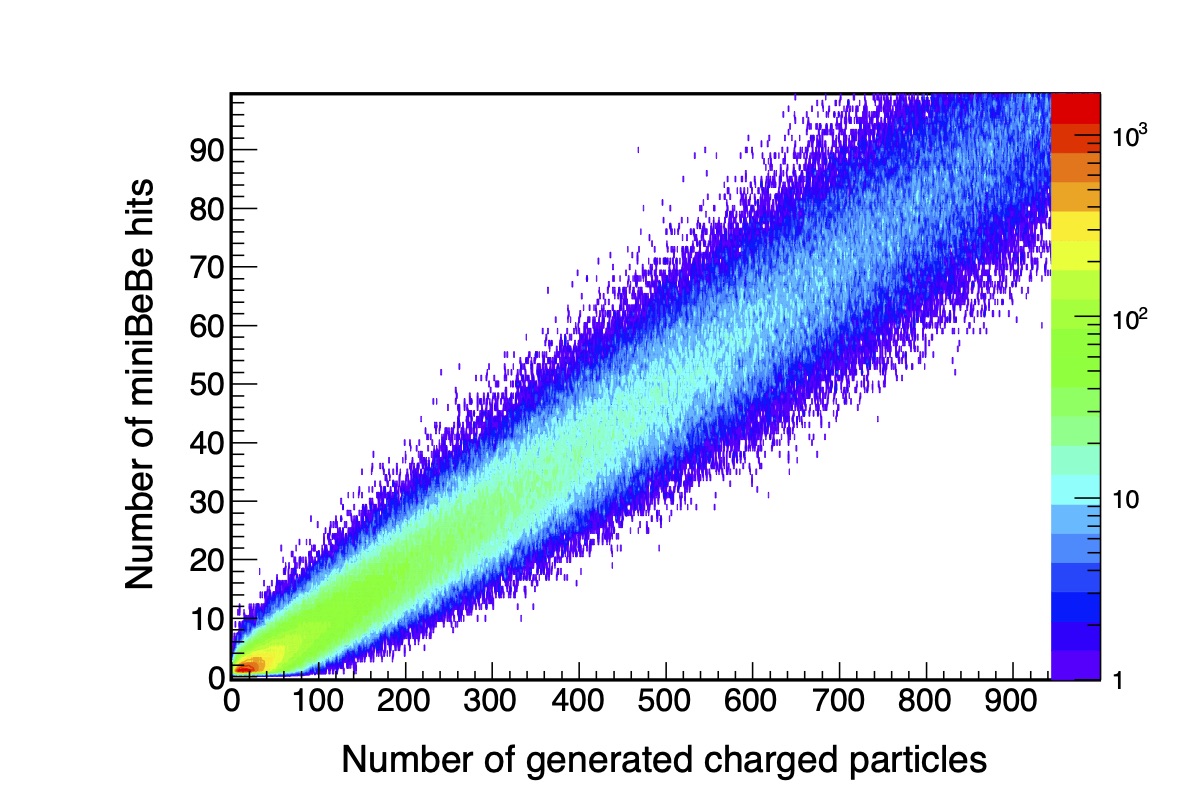}

\caption{Number of charged particles that hit the miniBeBe vs.\ the generated number of charged particles. The colors correspond to the total number of particles that hit the miniBeBe when produced by the simulation in a given charged particle multiplicity bin: blue corresponds to a small number whereas red corresponds to the largest number.}
\label{fig:minibebe-nch_multiplicity}
\end{figure}

\begin{figure}[!h]
\centering
\includegraphics[scale=0.2]{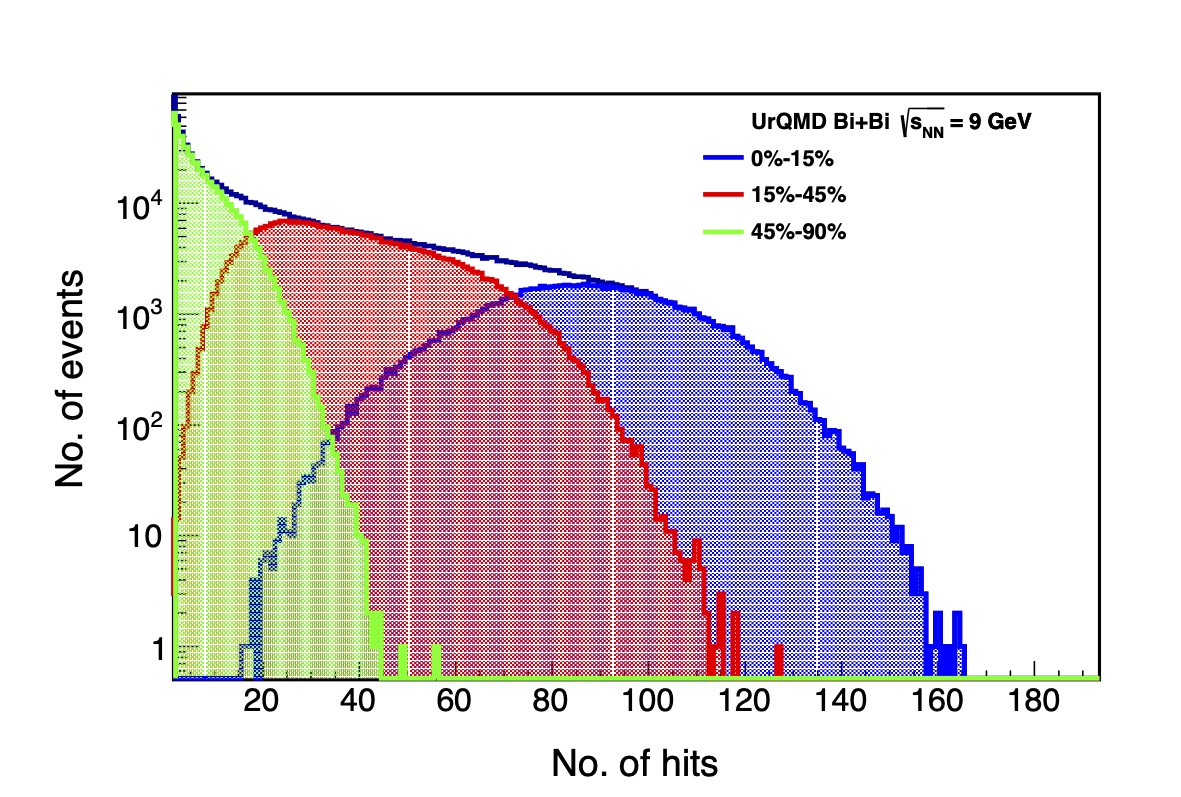}

\caption{MiniBeBe multiplicity per centrality range.}
\label{fig:minibebe_multiplicity}
\end{figure}

Figure \ref{fig:minibebe-nch_multiplicity} shows a (roughly linear) relation between the number of hits produced in the miniBeBe and the number of generated charged particles. This result is useful if we intend to produce an online centrality trigger with the miniBeBe. As shown in Fig.~\ref{fig:minibebe_multiplicity}, the miniBeBe raw multiplicity varies with respect to different centrality ranges. This behavior has been reported at higher energies in Ref.\ \cite{ALICE_centrality} where it is explained in terms of the geometrical properties of heavy ion collisions. Some events may be assigned to a wrong centrality range. This effect can be corrected offline during the data analysis or data reconstruction.

\subsection{Time information}\label{secVII_time}

\begin{figure}[!hbt]
\centering
\includegraphics[scale=0.2]{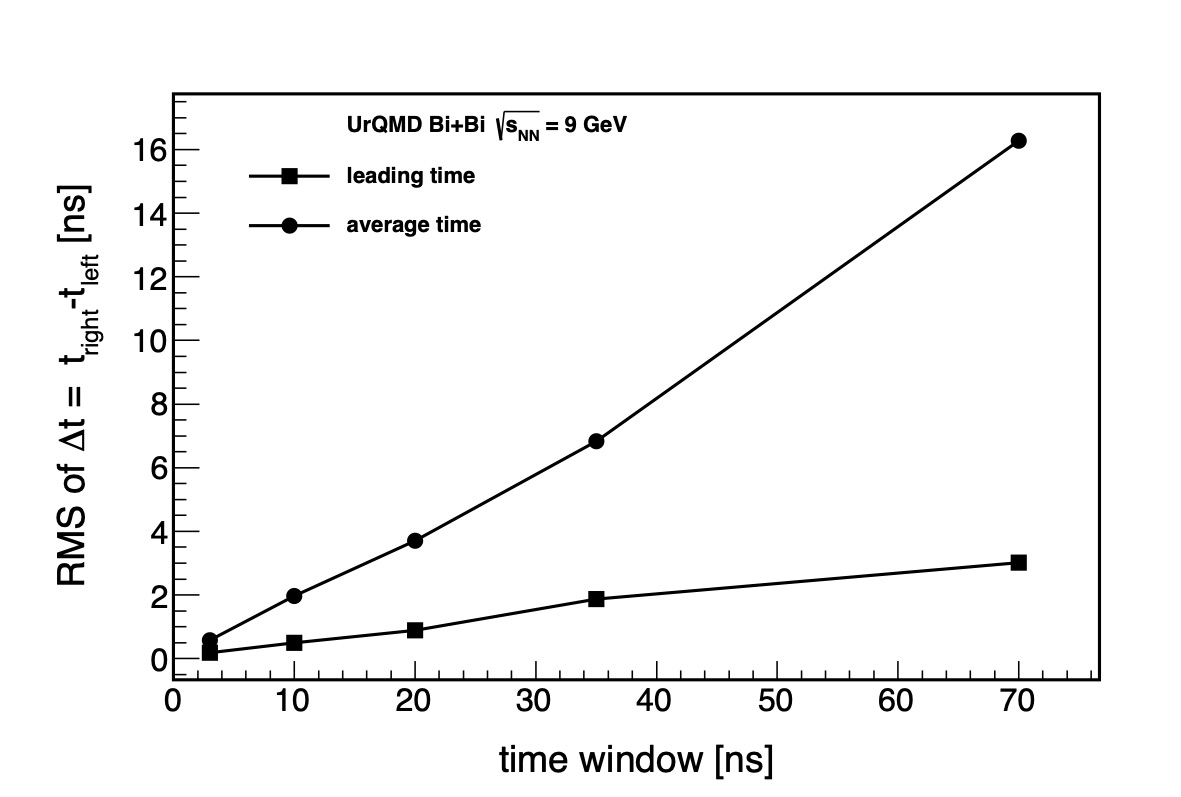}

\caption{MiniBeBe RMS time difference $t_{\rm right}-t_{\rm left}$ as a function of the time window.}
\label{fig:minibebe_time}
\end{figure}

The arrival time of the produced charged particles at individual cells was taken from the generated hit in MPDRoot. From the time information of the miniBeBe hits per event, we estimated the average hit time and the time-of-flight of the first charged particle reaching miniBeBe (leading time) for $z>0$, $t_{\rm right}$, and for $z<0$, $t_{\rm left}$. The root mean square (RMS) of the $\Delta t = t_{\rm right}-t_{\rm left}$ distribution provides an indication of the target for best time resolution of the miniBeBe. Figure~\ref{fig:minibebe_time} shows the RMS of the $\Delta t$ distribution as a function of several time windows: 3 ns, 10 ns, 20 ns, 35 ns and 70 ns where in each case we assumed that both the average and leading times, for $z>0$ and for $z<0$, are less than these time windows. 

\begin{figure}[!hbt]
\centering
\includegraphics[scale=0.2]{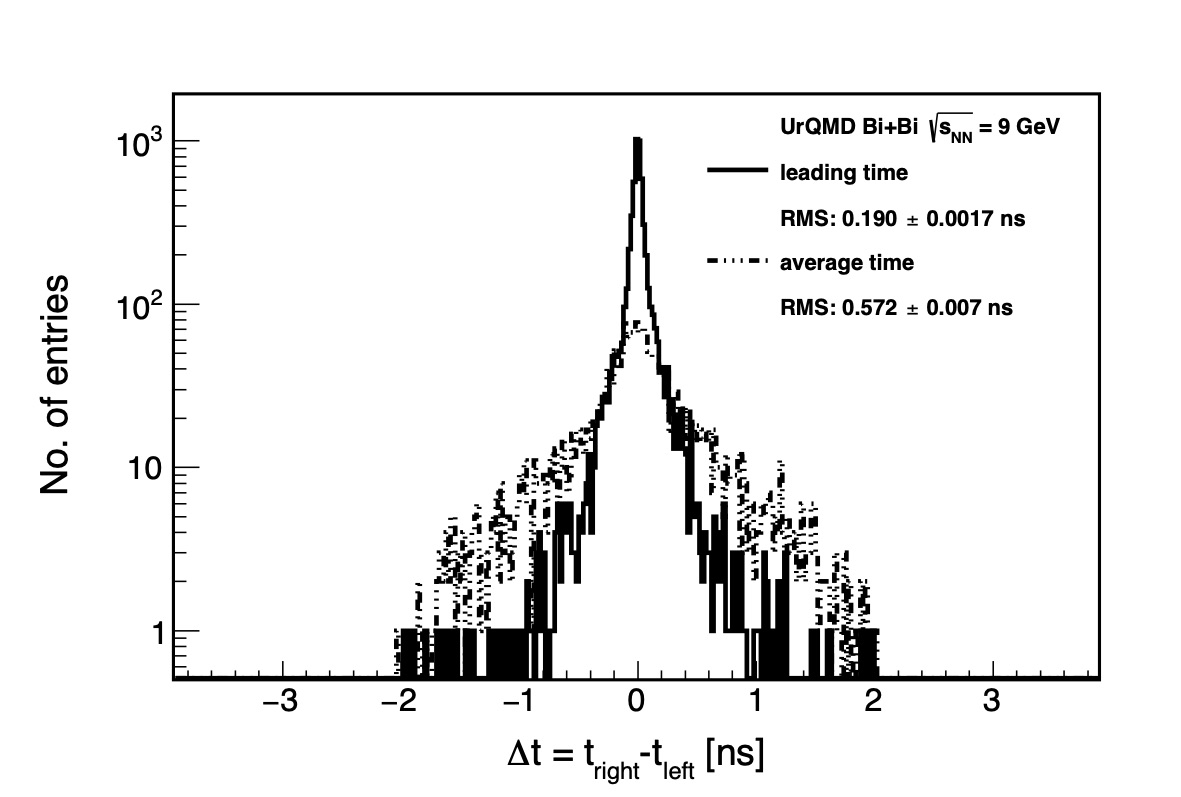}

\caption{Number of entries vs.\ $\Delta t$ for a time window of 3 ns calculated using the leading time (continuous line) and the average time (dotted line).}
\label{fig:minibebe_timedifference}
\end{figure}

As an example, Fig.~\ref{fig:minibebe_timedifference} shows the $\Delta t$ distribution for the average and leading time of the miniBeBe. It can be noted from Figs.~\ref{fig:minibebe_time} and ~\ref{fig:minibebe_timedifference} that the minimum RMS value for the $\Delta t$ distributions is obtained using the leading time for particles reaching miniBeBe. 

\begin{figure}[!hbt]
\centering
\includegraphics[scale=0.22]{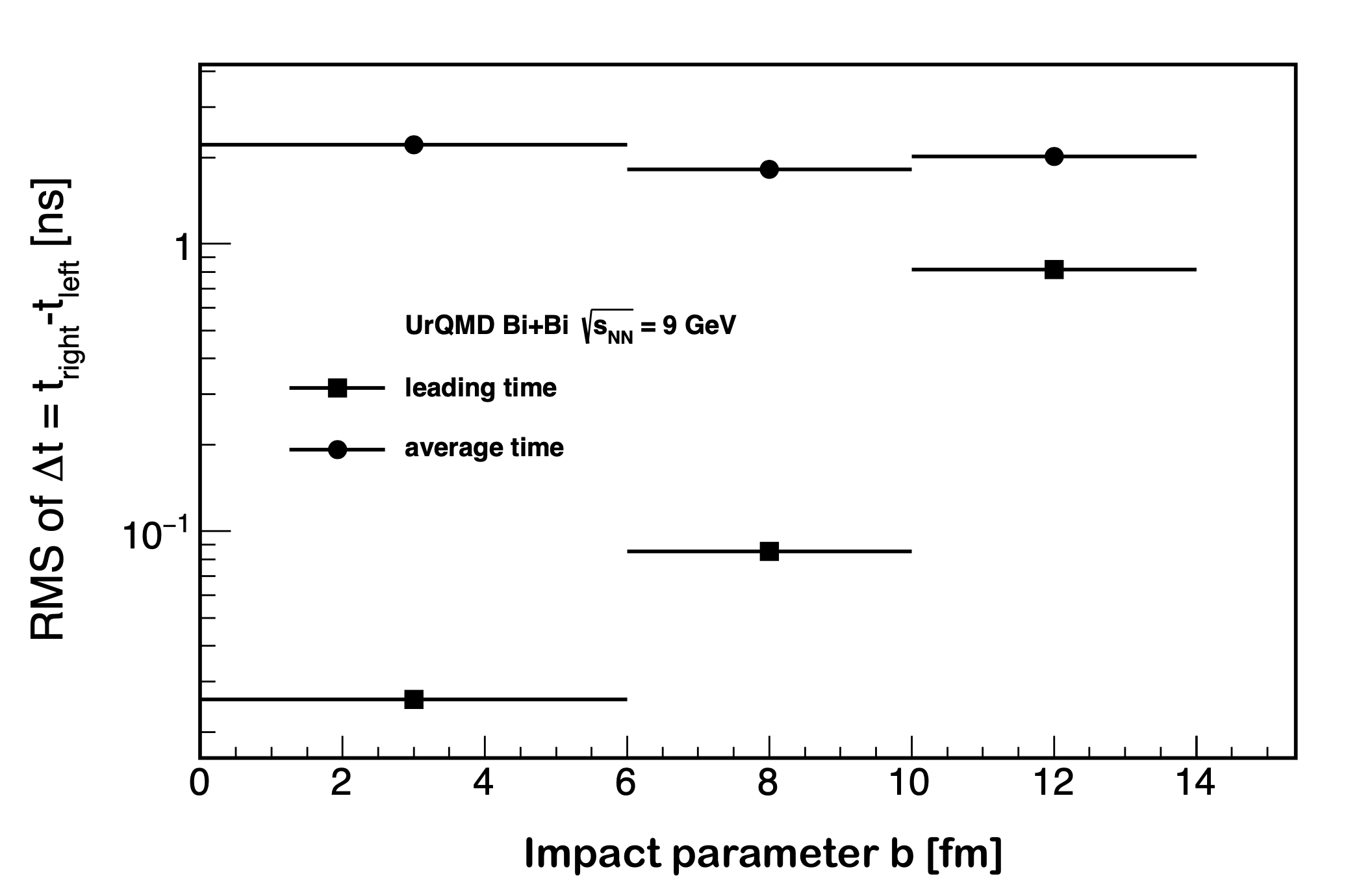}

\caption{RMS of the $\Delta t$ distribution as a function of the impact parameter $b$ of the collision.}
\label{fig:minibebe_timecentrality}
\end{figure}

The RMS value of the $\Delta t$ distribution depends also on the collision impact parameter $b$. The lowest RMS value of the $\Delta t$ distribution is obtained for central collisions, while for larger values of $b$ the RMS value is 0.815 ns, as can be seen in Fig.~\ref{fig:minibebe_timecentrality}. Thus, a time resolution of at least 0.026 ns is mandatory for the  miniBeBe to generate a proper beam-beam trigger signal based on the leading time measured by the miniBeBe data acquisition system. 

Using the leading time of miniBeBe hits, $t_{\rm right}$ and $t_{\rm left}$ for $z>0$ and $z<0$, respectively, we can determine with the miniBeBe the collision vertex along the $z$-axis as
\begin{center}
    $VertexMbb=\frac{t_{\rm right}-t_{\rm left}}{2} \times c$.
\end{center}

\begin{figure}[!hbt]
\centering
\includegraphics[scale=0.2]{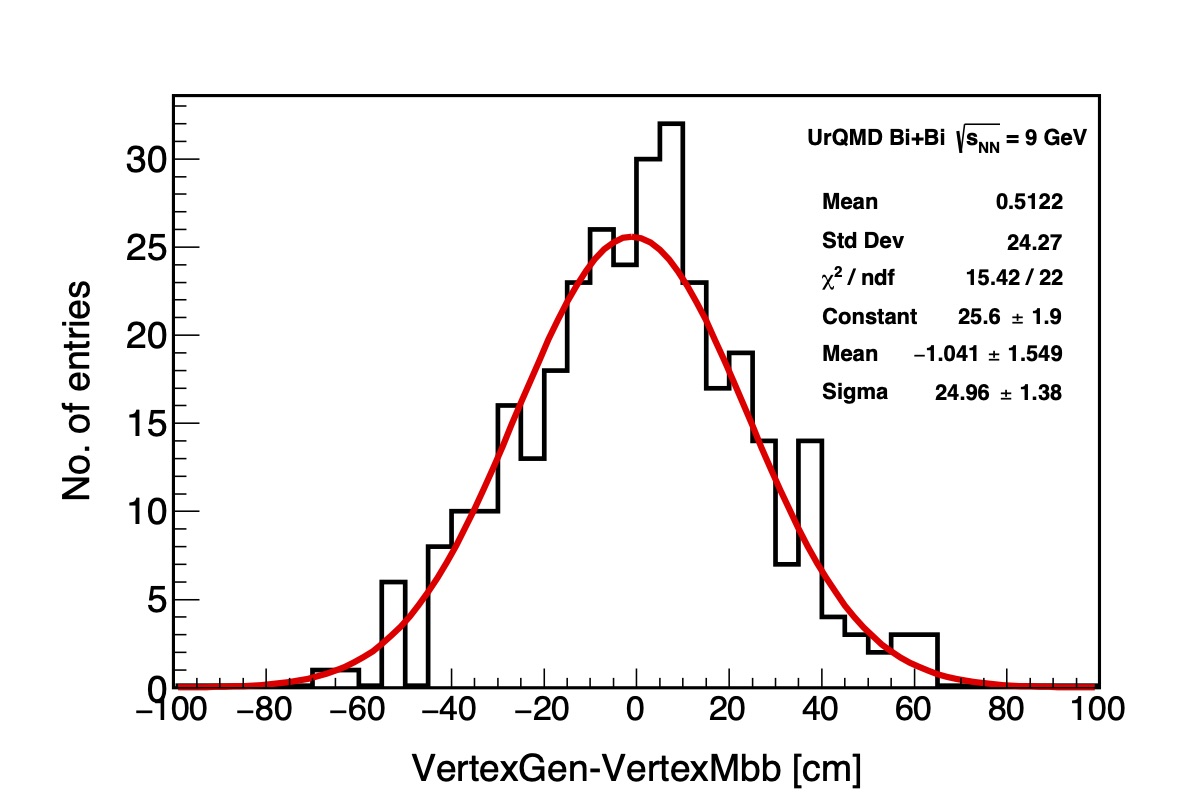}

\caption{The distribution of $\Delta vtx$, defined at the end of Sec. VIII C.
We show the difference between the generated vertex and the vertex determined with the leading time of the miniBeBe detector.}
\label{fig:minibebe_vtxreso}
\end{figure}

To estimate the resolution of the vertex determination of the miniBeBe, we computed the RMS of the $\Delta vtx=VertexGen-VertexMbb$ distribution, where $VertexGen$ is the generated position of the collision vertex given by the UrQMD generator.  Figure~\ref{fig:minibebe_vtxreso} shows the $\Delta vtx$ distribution. The best time resolution for the vertex determination using the miniBeBe is 24 cm/$c= 0.8$~ns.

\subsection{Beam-gas}\label{secVII_beamgas}

Beam-gas interactions are a background originated at a certain distance from the interaction point due to the interaction of the circulating particles in the beam with the residual gas in the beam pipe. This background depends on the NICA nominal bunch crossing. To simulate beam-gas events we generated p+O collisions with UrQMD at $\sqrt{s_{NN}}=9$ GeV with the collision vertex located at +19 m from the nominal interaction point, with a width of 3.5 m. 

\begin{figure}[hbt!]
\centering
\includegraphics[scale=0.2]{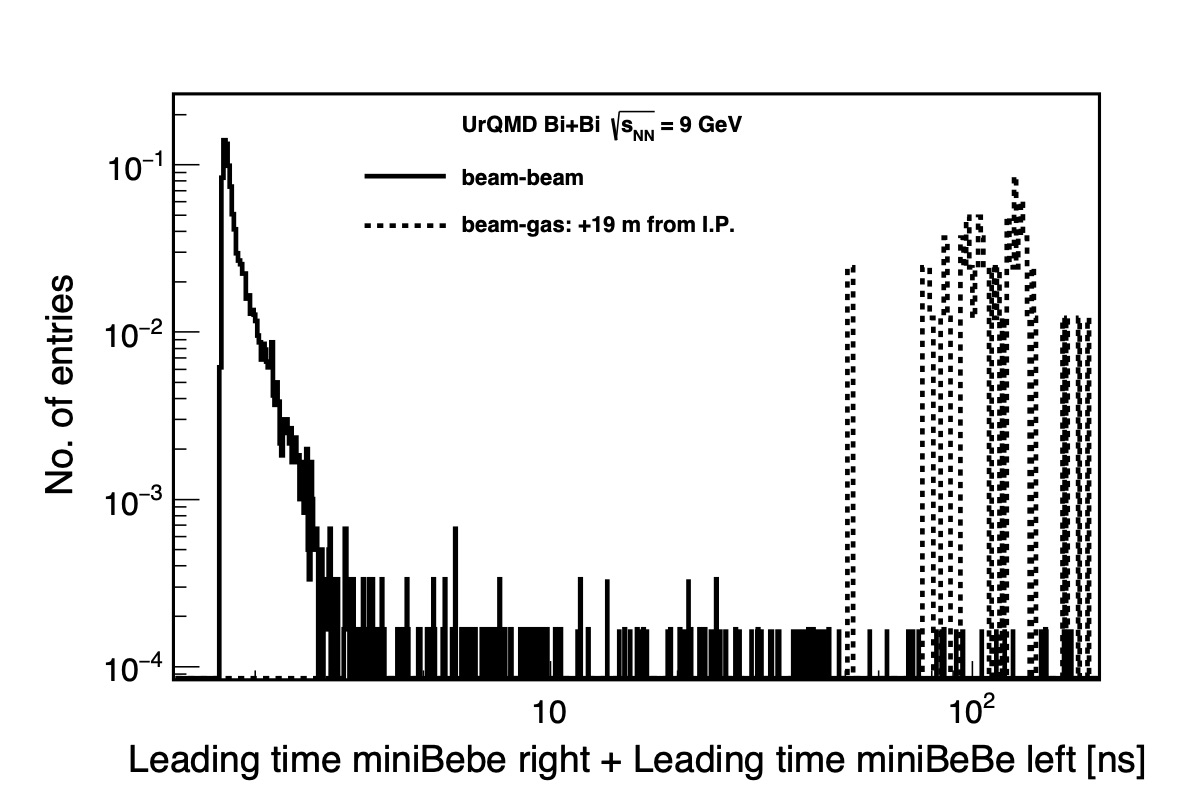}
\caption{The distribution of $t_{\rm right}+t_{\rm left}$. The sum of the leading time of the miniBeBe detector for $z>0$ and $z<0$ is shown for beam-beam and beam-gas generated events, 19~m from the interaction point.}
\label{fig:minibebe_beamgas}
\end{figure}

In order to evaluate the miniBeBe  capability to separate beam-gas interaction events from beam-beam collisions, we used the leading time distribution $t_{\rm right}+t_{\rm left}$ for beam-beam and beam-gas generated events. If the beam-gas interaction vertex events is located more than 19 m away from the interaction point, the miniBeBe may be able to discriminate beam-gas interactions from beam-beam collisions. Some beam-beam events at the tail of the $t_{\rm right}+t_{\rm left}$ distribution may be mistaken with beam-gas interactions and vice versa. As the location of beam-gas events is moved closer to the interaction point, the miniBeBe decreases its capability to veto beam-gas interactions, see Fig.~\ref{fig:minibebe_beamgas}. 
(No correction due to fine tuning cabling delay, neither time spread of the collision nor individual plastic scintillator cell time resolution, was applied to this analysis.)

\subsection{Summary of findings from MC simulations}\label{secVII_mbb_mpdroot_comments}

The results shown in this section can be summarized as follows:
\begin{itemize}
    \item The miniBeBe can generate a trigger signal for beam-beam collision events with a $\simeq 100$\% efficiency for the central rapidity region. For forward events, the trigger efficiency decreases below 80\%.
    \item The miniBeBe leading time is optimal to generate trigger signals, especially for high multiplicity events.
    \item A miniBeBe time resolution of 26 ps is needed to trigger central collision events. For non-central collisions, a not so stringent time resolution of only 85 ps is required. The miniBeBe will be able to provide a trigger signal with these requirements.
    \item The miniBeBe will be able to distinguish beam-gas interactions from beam-beam collisions if the vertex location of beam-gas events is far from the interaction point ($\gtrsim$ 19 m). If the location of  beam-gas vertex interactions is closer to the interaction point, the miniBeBe will become less efficient to set proper trigger flags to distinguish beam-beam from beam-gas events. 
    \item The number of hits in the miniBeBe seems to be sensitive to the centrality of the collision. This information may be useful to generate online centrality trigger classes. 
\end{itemize}

\section{\label{concl}Summary and Conclusions}

In this work we have presented the conceptual design for the miniBeBe detector that is proposed to be installed in the NICA-MPD to serve as a level-0 trigger for the TOF. We have described the detector sensitive elements and the read-out electronics. We have performed simulations to show that the design is capable to provide an efficient trigger for low and high multiplicity events. The miniBeBe capabilities to additionally serve as a beam-gas veto as well as to determine the beam-beam vertex are also shown. The prototype of some of its parts is currently being developed and will soon be tested in a radiation hard environment. 

It is important to mention that, as it usually happens with any other detector concept, the current design is evolving to better suit the needs of the MPD as a whole. These needs are now being discussed within the MPD Collaboration which may result in a scaling up of the design. The conceivable modifications include a larger longitudinal dimension, a smaller radius as well as an increase of the number of sensitive elements in the azimuthal direction. Nevertheless, it is important to bear in mind that all the simulations that were performed for the dimensions hereby discussed still stand and that a larger number of sensitive elements can only increase the detector capabilities. Also, the mechanical integration with the support is being actively explored as well as the integration with other MPD subsystems. Moreover, some of the capabilities of the miniBeBe could be enhanced if used together with the BeBe detector that we have also proposed to be considered as a forward beam-beam counter~\cite{BeBe}. The technical design for the detector will be reported in a more detailed document elsewhere.

\vspace*{-2mm}
\section*{Acknowledgements}
\vspace*{-1mm}

The authors thank Adam Kisiel, Marcin Bielewicz, Slava Golovatyuk and Itzhak Tserruya for very useful comments and suggestions. We also acknowledge the MPD Collaboration for facilitating the use of the MPD\-Root framework to carry out the simulations and for the continuous interaction and feedback. I.M. thanks the ICN-UNAM faculty and staff for support and kind hospitality provided during 
part of this work. L.R. thanks the BUAP Medical Physics and Elementary Particles Labs and their faculty for their kind hospitality and support during 
part of this work. The authors are in debt to Luciano D\'iaz and Enrique Murrieta for their technical support. Financial support for this work has been received from UNAM-DGAPA-PAPIIT grant number IG100219 and from the Consejo Nacional de Ciencia y Tecnolog\'ia (CONACyT), grant numbers A1-S-7655 and A1-S-16215. I.M. acknowledges support from a postdoctoral fellowship granted by CONACyT. M.R.C. thankfully acknowledges the permission to use computer resources, the technical advise and the support provided by the Laboratorio Nacional de Superc\'omputo del Sureste de M\'exico (LNS), a member of the CONACyT national network of laboratories, with resources from grant number 53/2017.



\end{document}